\documentclass[twocolumn]{aastex6}

  \usepackage{amsmath}
  \usepackage{tabularx}
  \usepackage{color}
  \usepackage{graphicx}
  \usepackage{rotating}
\usepackage{longtable}

\usepackage{filecontents}

\begin{document}

\title{High redshift BL Lac objects: spectroscopy of candidates \\
     }

\author{M. Landoni\altaffilmark{1},  S. Paiano\altaffilmark{2,3}, R. Falomo\altaffilmark{2}, R. Scarpa \altaffilmark{4}, A. Treves\altaffilmark{5}}

\altaffiltext{1}{INAF, Osservatorio Astronomico di Brera, Via E. Bianchi 46 I-23807 Merate (LC) - ITALY}
\altaffiltext{2}{INAF, Osservatorio Astronomico di Padova, Vicolo dell'Osservatorio 5 I-35122 Padova (PD) - ITALY}
\altaffiltext{3}{Universit\`a di Padova and INFN, Via Marzolo 8, I-35131 Padova - ITALY}
\altaffiltext{4}{Instituto de Astrofisica de Canarias, C/O Via Lactea, s/n E38205 - La Laguna (Tenerife) - ESPANA}
\altaffiltext{5}{Universit\`a degli Studi dell'Insubria, Via Valleggio 11 I-22100 Como - ITALY}

\begin{abstract}
We report on 16 BL Lacertae objects that were proposed to be at $z > 1$. 
We present spectroscopic observations secured at the 10.4m GTC that allowed us to assess the redshift of these sources. In particular, for five objects we disprove the previous value of the redshift reported in literature and found that they lie at $z < 1$. Moreover, two of them exhibit broad emission lines that are not characteristic of BL Lacertae object. On the other hand, for eight targets we improve the tentative value of $z$, previously based on only one feature, by detecting a number emission lines. Finally, in three cases we detect onset of Ly-$\alpha$ forest at $z > 2.50$. Based on the new high quality spectra we found that only half of the observed objects can be classified as bona-fine BL Lacs.

\end{abstract}


\section{Introduction} \label{sec:intro}

BL Lac objects (BLLs) are active galactic nuclei that show a strong, luminous continuum that arises from non-thermal emission and are characterised by  variability at all wavelengths, from near-IR to $\gamma$-rays (see e.g. \cite{falomo2014}, \cite{madejski2016}). These properties are explained assuming the presence of a relativistic jet pointed in the direction of the observer. BLLs, which dominate the extragalactic $\gamma$-rays sky, are a unique laboratory for high energy astrophysics and fundamental physics. However the determination of their distance, which is crucial to fully understand their physical properties, is a challenging task. In fact their optical spectrum is quasi-featureless and the non-thermal continuum outshines both the superposed thermal contribution due to the stellar component of the host galaxy and the emission lines generated by fluorescence in clouds surrounding the central black hole, thus preventing to determine the redshift. Moreover, when spectral lines are detected they are characterised by very small (few $\textrm{\AA}$) equivalent width (EW) and thus high quality optical spectra in terms of Signal to Noise Ratio (SNR) and spectral resolution are required to reveal them (e.g. \cite{sbarufatti2006}, \cite{shaw13}, \cite{pita14}, \cite{landoni2014}, \cite{paiano2018,paiano2017b,paiano2017}). 
\\

Up to now only very few BLLs objects have a well determined redshift at $z \gtrsim 1$. For instance, according to the latest version of the Roma BZ-Catalog \citep{massaroe15, massaroref} there are $\sim$ 1000 sources classified as genuine BLL objects and for about 600 targets the redshift is unknown.  For the remaining 400 there is a value for the redshift (but for $\sim$ 100 the $z$ is tentative) and only 15 objects are reported at $z$ $>$ 1.  Remarkably,  for 12 of them the value of $z$ is also uncertain.  The discovery of a number of BLL objects at $z \gtrsim 1$ is fundamental to assess their luminosity function and cosmic evolution, in particular to clarify if it is positive or negative (e.g. \cite{ajello14} and references therein). Moreover, when detected at very high energy with Cherenkov Telescopes, BLLs at $z \gtrsim 1$ are precious probes to constrain the Extragalactic Background Light \citep[e.g.][]{franceschini17} and also to test some exotic effects coming from non-standard physics, like Lorentz-Invariance Violation \citep[e.g.][]{tavecchio16} or to indirectly detect Axion Like Particles \citep[e.g.][]{deangelis}. 
\\

Here we focus on a sample of BLLs proposed to be at $z > 1$ and selected from the Sloan Digital Sky Survey (SDSS), see next section.
\\
The paper is organised as follows: we give details on sample selection and data reduction in Section 2 while we present our main results in Section 3. Notes on individual observed sources are reported in Section 4 and we draw our conclusions in Section 5. Throughout  the paper we adopt the following cosmological parameters: H$_0=$ 70 km s$^{-1}$ Mpc$^{-1}$, $\Omega_{\Lambda}$=0.7, and $\Omega_{m}$=0.3.

\section{Sample, reduction and data analysis} \label{sec:sample}
We selected targets that are candidates for being at high redshift ($z \gtrsim 1$) from the list of BLLs sources, based on SDSS Data Release 7 spectra provided by \cite{plotkin10} (see Table 4 in their paper). This dataset includes 637 objects and only half of them (367) have redshift . On the latter, only 28 posses indication that $z > 1$ and for 18 the measurement is also uncertain. We have undertaken a spectroscopic study of these targets using high SNR spectra secured at large telescopes. From the initial database, we excluded objects with apparent magnitude $\gtrsim 20$, yielding a sample of 24 sources. In this paper we present 16 objects observed with Gran Telescopio CANARIAS (GTC) equipped with OSIRIS. Details are reported in Table 1. 
\\

The observations were obtained in service mode at the GTC using the low resolution spectrograph OSIRIS \citep{cepa2003}. 
The instrument was configured with the grisms R1000B and R1000R\footnote{http://www.gtc.iac.es/instruments/osiris/osiris.php}, in order to cover the whole spectral range 4100-9500~$\textrm{\AA}$, and with a slit width~$=$~1'' yielding a spectral resolution $\lambda$/$\Delta\lambda$~$=$~800.
\\
For each grating we obtained three individual exposures sequentially under the same sky condition and instrument configuration that were combined in order to perform a good cleaning of cosmic rays and CCD cosmetic defects. Wavelength calibration was performed using the spectra of Hg, Ar, Ne, and Xe lamps and provide an accuracy of 0.1~$\textrm{\AA}$ over the whole spectral range. For each object the spectra obtained with the two grisms were joined into a final spectrum covering the whole spectral range.
Data were corrected for atmospheric extinction using the mean La Palma site extinction table\footnote{https://www.ing.iac.es/Astronomy/observing/manuals/}.  Relative flux calibration was provided by spectro-photometric standard stars secured during the same night of the target observation. Absolute flux calibration was assessed using relative photometry of objects in the Sloan r filter acquisition image. The accuracy of the calibration is about $\Delta m \sim 0.1$.
We exploited for our analyses the software procedures developed for our long-term optical spectroscopy programs at large telescopes (like ESO-VLT and GTC) specifically aimed to measure the redshifts of BLLs (\cite{sbarufatti2006}, \cite{sandrinelli2013}, \cite{landoni2015}, \cite{paiano2016}, \cite{paiano2017b,paiano2017}). Detailed information on the observations and target magnitudes at time of data acquisition are given in Table 2.

\section{Results} \label{sec:results}
We report the spectra of BLLs 
in Fig. 1, close up of the faintest spectral features in Fig. 2 and we provide the full figure set and machine readable data online at \url{http://archive.oapd.inaf.it/zbllac/}.
We enhance spectral features by showing both the flux calibrated and the normalised spectrum that was obtained by dividing the calibrated
spectrum by a power-law continuum in all cases where the spectrum could be fitted in this way (see also individual notes in Section 4). This procedure allows us to emphasise weak features that appears clearer on the normalised data.  
\begin{figure*}
	\centering
	\includegraphics[width=12cm, angle=270]{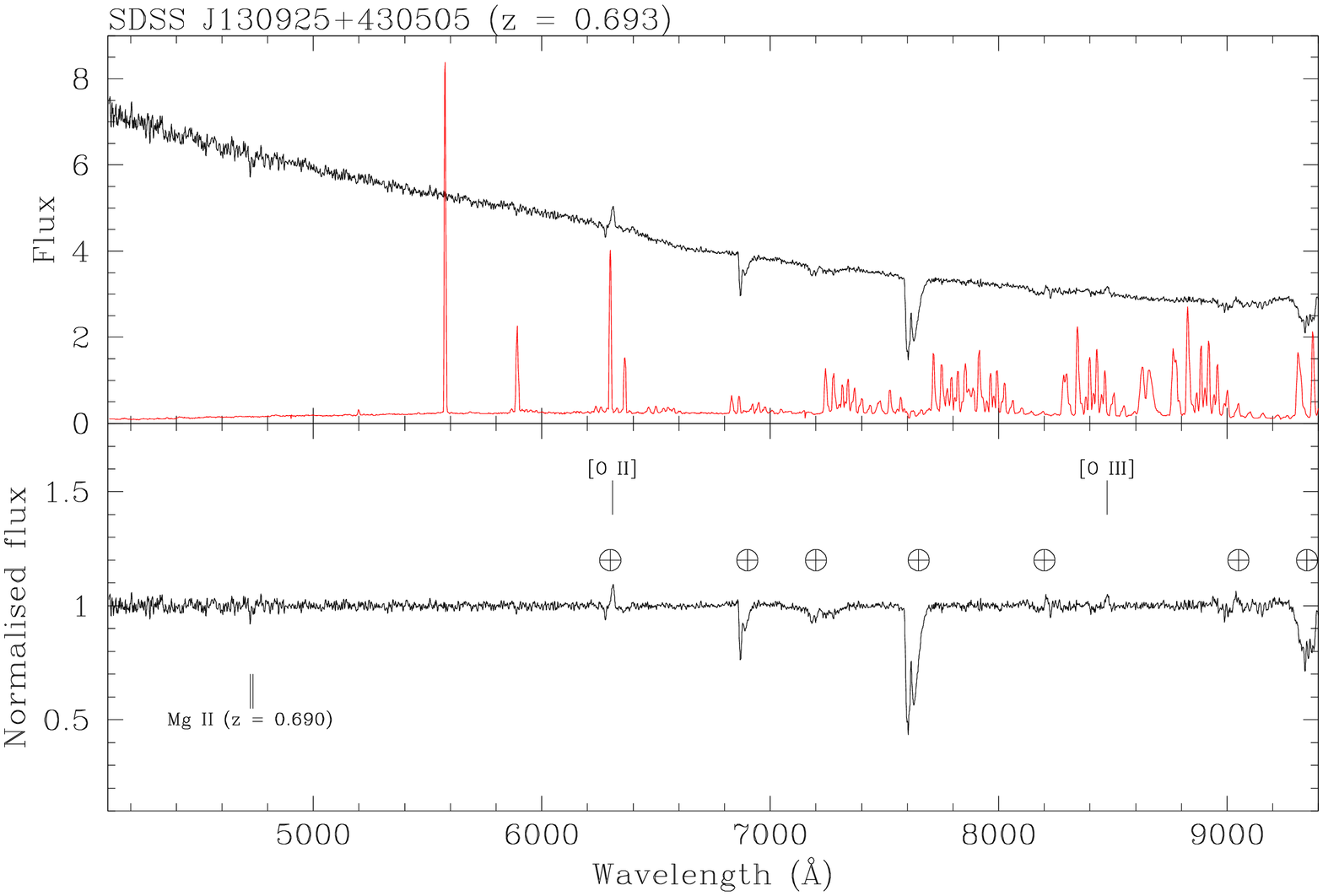}
    \caption{Sample spectrum of the high redshift SDSS BLLs obtained at GTC. \textit{Top panel}: Flux calibrated and deredded spectra in black, while the red solid line is the sky spectrum in arbitrary units reported for sake of comparison. \textit{Bottom panel}: Normalized spectra. The main telluric bands are indicated by $\oplus$, the absorption features from interstellar medium of our galaxies are labelled as IS (Inter-Stellar). Data available at URL: http://www.oapd.inaf.it/zbllac/}
    \label{fig:example_pair}
\end{figure*}

We evaluate
the signal-to-noise ratio (S/N) in a number of 
spectral regions (see Table 3) adopting the procedure described in \cite{paiano2017}. The use of GTC+OSIRIS enabled us to obtain spectra with SNR superior to that of SDSS (about a factor of $\sim 2$). In details,  the average S/N of GTC spectra is assessed to be $\sim$ 35 while for SDDS data is $\sim$ 15 and no significant variabilities of the targets in terms of apparent r band magnitude have been detected.
All spectra were carefully inspected to search for emission
and or absorption lines. When a possible feature is found we check its reliability by comparing the three independent exposures (see Section 2) and we consider it reliable if detected above the noise level in all the three acquired frames.
\\

For 5 objects, we disprove the value of the redshift previously reported in literature. For one of them, SDSS J092902.42$+$194525.1 the spectrum appears completely featureless and we infer z $\geq 0.35$ from the absence of starlight from host galaxies \citep{sbarufatti2006,paiano2017b} while for the remaining sources we give new values for the redshift. In particular, for SDSS J072659.52$+$373423.0 we measure z = 0.791 by detecting Ca II H-K lines (3934-3968 $\textrm{\AA})$ from the host galaxy while for SDSS J101115.63$+$010642.5 we found at $z = 0.857$ emission features from [Ne V] (3346 $\textrm{\AA}$), \mbox{[O II] (3727 $\textrm{\AA}$),} H$\beta$ (4861 $\textrm{\AA}$) and [O III] (4959-5007 $\textrm{\AA}$). In the case of SDSS J124700.72+442318.8 we confirm the presence of a feature at $\lambda$ 7860 $\textrm{\AA}$ but the identification with Mg II (2800 $\textrm{\AA}$), as suggested by \cite{plotkin10}, is unlikely since it appears rather narrow (FWHM $\sim$ 300 km s$^{-1}$) and weak (EW $\sim 2.30 \textrm{\AA}$) and we suggest identification as [O III] (5007 $\textrm{\AA}$) at z = 0.569. Finally, for SDSS J132802.09$+$112913.6 we detect emission lines from Mg II (2800 $\textrm{\AA}$), [Ne V] (3346 $\textrm{\AA}$) and [O II] (5007 $\textrm{\AA}$) all at the same redshift \mbox{$z = 0.580$}.
\\

About the remaining 11 targets, for three of them (namely SDSS J120059.69$+$400913.1, SDSS J123132.37$+$013814.1 SDSS J145059.98$+$520111.7) a spectroscopic lower limit to $z$ is set on the basis of detection of onset of the Ly-$\alpha$ forest. We confirm the literature values and we did not find any further emission lines redward Ly-$\alpha$ forest although our new GTC data present in that regions a SNR a factor of $\sim$ 5 higher than SDSS spectra. For eight cases (SDSS J090818.95+214820.0, SDSS J094257.81-004705.2, SDSS J121037.35+525341.9, SDSS J124030.93+344527.5, SDSS J130925.52+430505.5, SDSS 144050.14+333350.2, SDSS J152422.56+374034.1 and SDSS J170108.89+395443.0), we endorse the proposed tentative value of the redshift that was based on only one emission line, by measuring further spectral features that enforce the values of $z$. We report in Table 4 all the spectral features identified in our sample and discuss each target in detail in the next section. 
\section{Notes for individual sources } \label{sec:notes}
\setcounter{figure}{2}
\begin{figure}

	\centering
	\includegraphics[width=9cm]{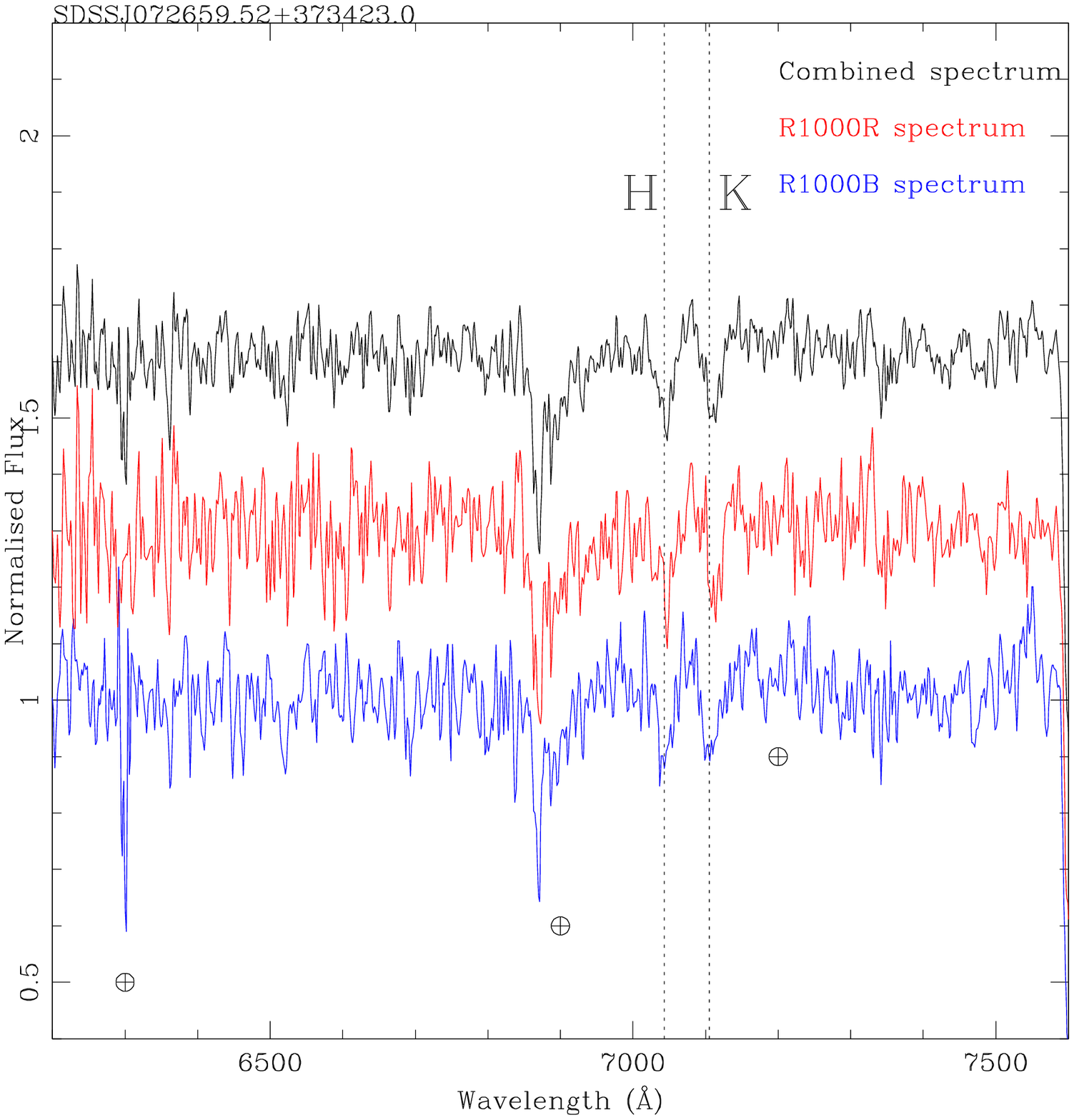}
    \caption{Close up of the detected H-K band of Ca II in the spectrum of SDSS J072659.52+373423.0. Black line: Joint R1000B+R1000R grism secured spectra. Red line: R1000R spectrum; Blue line: R1000B spectrum. See Table 2 for details on observations}
    \label{fig:example_pair}
\end{figure}
\label{individual}

\begin{itemize}
\item[] \textbf{SDSS J072659.52+373423.0}:
On the basis of emission feature at $\lambda$ 7200 $\textrm{\AA}$ in the SDSS spectrum \cite{plotkin10} suggested a tentative redshift z = 1.577. We do not confirm the presence of this feature in our spectrum (SNR = 20) while we detect an absorption doublet at $\lambda \lambda$ 7043-7107 $\textrm{\AA}$ that we identify as Ca II (3934-3968 $\textrm{\AA}$) at $z = 0.791$. This feature, although it is encompassed by two atmospheric absorption bands, is firmly detected since it is revealed in both R1000B and R1000R spectra (see Figure 3). 

\item[] \textbf{SDSS J090818.95+214820.0}:
A redshift $z = 2.089$ was proposed by \cite{plotkin10} on the basis of a weak emission line at 8600 $\textrm{\AA}$ ascribed to Mg II (2800$\textrm{\AA}$). We confirm the presence of this emission line at 8650 $\textrm{\AA}$ in our data (SNR = 30) and we also detect a broad feature at $\sim$ 4789$\textrm{\AA}$ corresponding to C IV (1909 $\textrm{\AA}$) therefore firmly confirming the redshift.

\item[] \textbf{SDSS J092902.42+194525.1}:
\cite{plotkin10} proposed a tentative redshift $z = 1.774$ based on a narrow emission line at 7770 $\textrm{\AA}$. We do not confirm this structure and suspect that it might be an arctifact due to filtering procedure of SDSS spectrum since it is not present in the unfiltered data. No other emission or absorption lines are detected in our spectrum (SNR = 25), therefore the redshift remains unknown. From the minimum detectable EW$_{min}$ of starlight features from the host galaxy \citep{paiano2017} we set a lower limit to $z \geq 0.35$. According to the expected number of Mg II intervening systems for sources at $z > 1$, the probability of not detecting any one of them as in the case of this object is roughly 0.25 suggesting that it is unlikely that this BLL lies at high redshift.

\item[] \textbf{SDSS J094257.81-004705.2}:
In our optical spectrum (SNR = 35) we detect two emission lines at 4509 $\textrm{\AA}$ and 6610 $\textrm{\AA}$  that we identify as CIII] and Mg II at $z = 1.36$. This confirms the previous tentative redshift based only on one spectral feature \citep{plotkin10}. In addition, we also measure an intervening Mg II absorption system at \mbox{z = 0.819}.

\item[] \textbf{SDSS J101115.63+010642.5}:
A single broad emission feature at $\sim$ 9200 $\textrm{\AA}$  is visible in the SDSS spectrum. It was erroneously interpreted as Ly-$\alpha$ by SDSS automatic procedure leading to the tentative redshift $z \textgreater 5$. In the work of \cite{plotkin10} the authors suggest a tentative redshift of z = 1.479 on the basis of narrow feature at $\sim$ 6930 $\textrm{\AA}$.  In our new data (SNR = 50) we detect the two features at 6929 $\textrm{\AA}$ and 9209 $\textrm{\AA}$ which we interpret as [O II] and [O III] doublet. Moreover, other fainter emission lines due to [Ne V]  and H$\beta$ that are all consistent at more modest redshift $z = 0.857$ are revealed.

\item[] \textbf{SDSSJ120059.69+400913.1}:

The lower limit of redshift of the source $z \geq 3.37$ was derived robustly by the SDSS and confirmed in the sample of \cite{plotkin10} by detecting firmly the onset of the Ly-$\alpha$ forest. Our new GTC data (SNR = 25) presents a SNR about a factor of $\sim$ 5 higher than SDSS in the region 5400-9200$\textrm{\AA}$ and, apart from absorption features from intervening systems, does not present any emission line down to EW $\geq 0.70$ .

\item[] \textbf{SDSS J121037.35+525341.9}:
A faint broad emission line at 5350 $\textrm{\AA}$  is visible in the SDSS spectrum. If identified as MgII (2800 $\textrm{\AA}$) a tentative redshift $z = 0.916$ is derived \citep{plotkin10}. The newly GTC-OSIRIS data (SNR = 40) confirm the presence of the feature at same wavelength and in addition we detect clearly the [O II] (3727 $\textrm{\AA}$) emission line at $\lambda$ 7142 $\textrm{\AA}$.

\item[] \textbf{SDSS J123132.37+013814.1}:
The lower limit of the redshift was set $z \geq 3.147$ from the detection of the Ly-$\alpha$ forest in the SDSS spectra. We confirm the detection of the Ly-$\alpha$ forest \citep{plotkin10} at the same redshift while we do not measure any further emission line in the region 5200-9200 $\textrm{\AA}$ at (SNR = 30) and with EW $\geq \sim 0.40$ making this source, apart from line-of-sight intervening systems, intrinsically featureless.

\item[] \textbf{SDSS J124030.93+344527.5}:
A tentative redshift of $z = 1.642$ was proposed on the basis of a marginally detected weak emission line at  $\lambda \sim$ 7390 $\textrm{\AA}$ ascribed to Mg II (see \cite{plotkin10}). We confirm the presence of the feature and we further detect an intervening absorption system at $z = 1.449$ by measuring absorption lines from Mg II and Fe II all the same redshift. We also note the presence of narrow absorption features in the GTC spectrum (SNR = 50) whose identification is not obvious and could be potentially associated to Fe II intervening systems at $z \leq 1.45$. The continuum shape is more complex than usual BL Lac sources and cannot be described by a single power-law in the optical region. We thus adopted a polynomial function to assess a fit of the continuum used to emphasise faint spectral line.

\item[] \textbf{SDSS J124700.72+442318.8}:
The only available spectroscopical redshift estimation of  z = 1.812 comes from the SDSS Sky Survey and it is based on a narrow emission line at $\lambda$ 7857 $\textrm{\AA}$ ascribed to Mg II (2800 $\textrm{\AA}$). We confirm the alleged emission lines in our spectrum (SNR = 60) at the same wavelength, however the proposed interpretation as Mg II appears dubious. A more likely interpretation is emission from \mbox{[O III] (5007 $\textrm{\AA}$)} at $z = 0.569$ on the basis of the very small FWHM ($\sim$ 300 km s$^{-1}$) and EW ($2.30 \textrm{\AA}$) . The target is also detected in the $\gamma$-ray band by FERMI satellite (Acero et. al 2016).

\item[] \textbf{SDSS J130925.52+430505.5}:
\cite{plotkin10} proposed a tentative redshift of $z = 1.154$ but no further indications of spectral features that sustain the measurements are reported. The target has been observed with the Keck telescope (equipped with LRIS) by Shaw et al 2013, where they propose $z = 0.69$ interpreting the narrow emission line at $\sim$ 6298 $\textrm{\AA}$ as [O II] (3727 $\textrm{\AA}$). In our higher SNR spectrum (SNR = 100), we confirm the presence of the [O II]  (3727 $\textrm{\AA}$) and we also clearly detect the emission line from [O III] (5007 $\textrm{\AA}$) consistently at z = 0.693.  In addition we also detect an absorption feature that we ascribe to Mg II system at $z = 0.690$ probably associated to the cool gaseous halo of the BLL. We remark also that this object belongs to the detected sources in FERMI 3LAT catalog (Acero et. al. 2016).

\item[] \textbf{SDSS J132802.09+112913.6}:
\cite{plotkin10}, based on a single weak emission line detected at 7915 $\textrm{\AA}$ in the SDSS spectrum  and interpreted as Mg II (2800 $\textrm{\AA}$), proposed a tentative redshift $z = 1.827$.  This value is not confirmed by our data (SNR = 30) since we clearly detect two prominent broad emission lines at 4409 $\textrm{\AA}$ (EW = 86 $\textrm{\AA}$) ascribed to Mg II and [Ne V]  consistently at $z = 0.580$. The feature at 7915 $\textrm{\AA}$ is indeed due to [O III] (5700 $\textrm{\AA}$) (see Figure 1). 

\item[] \textbf{SDSS J144050.14+333350.2}:
From a single broad emission line heavly contamined by strong telluric band of $O_2$, a tentative redshift $z = 1.774$ was suggested by \cite{plotkin10} interpreting it as Mg II. In addition to this feature, we also find further emission lines from C IV (1550 $\textrm{\AA}$) and C III] (1909 $\textrm{\AA}$) at redshift $z = 1.747$ that is slightly different than the previously suggested one.  In fact, the broad feature identified by \cite{plotkin10} and detected also in our data corresponds to transition of Mg II but with its blue wing strongly depressed by telluric lines. Finally, we also detect in our new data (SNR = 30) two intervening Mg II absorption systems at z = 0.578 and z = 1.168.

\item[] \textbf{SDSS J145059.98+520111.7}:
The source was already discussed in \cite{paiano2017b}.

\item[] \textbf{SDSS J152422.56+374034.1}:
On the basis of a tentative emission feature at $6200 \textrm{\AA}$ \cite{plotkin10} proposed redshift of z = 1.219. In our high quality spectrum (SNR = 40), we confirm the detection of this broad emission line at $\lambda$ 6204 $\textrm{\AA}$ ascribed to Mg II (2800 $\textrm{\AA}$) and, moreover, an emission from C III] (1909 $\textrm{\AA}$) at $\lambda$ 4235 at the same redshift.

\item[] \textbf{SDSS J170108.89+395443.0}:
We detect firmly (SNR = 25) the emission line from Mg II (2800 $\textrm{\AA}$) and C IV (1550 $\textrm{\AA}$) at the redshift z = 1.895. This confirms the tentative value proposed by \cite{plotkin10}.

\end{itemize}

\section{Conclusions}

We obtained high quality optical spectra for 16 objects proposed to be BLLs at z $\geq$ 1. For five of them we disprove the previous value of the redshift and find that they are indeed at z $<$ 1. For 8 other targets, for which the redshift was based on only one spectral line, we are able to establish the value of $z$ by detecting additional features. 
According to their spectral characteristic, we found that only seven sources of the sample are bona fide BLLs, while for the remaining six broad emission lines are revealed indicating that they should be re-classified as Flat Spectrum Radio Quasars.
\\

Finally, we comment on the remaining 3 sources where no emission lines are detected. In these targets only intervening absorption systems are apparent and we confirm the lower limit of $z \geq 3.367, \geq 3.140, \geq 2.470$ for SDSS J120059.69+400913.1, SDSS J123132.37+013814.1, and SDSS J145059.98+520111.7 respectively.  In these three spectra no emission lines with EW $>$ 1 $\textrm{\AA}$ are detected . This implies that if emission lines are present at redshift close to the highest z intervening system, their luminosity should be less than $\sim$ $10 \times 10^{42}$ erg s$^{-1}$ . These values are much smaller than the typical distribution of emission line luminosity found for QSOs \citep{shen11}. This supports the fact that these targets are true BLL objects. They are the most distant BLLs ever detected and remarkably one of them (SDSS J145059.98+520111.7) is also $\gamma-$ray emitter detected by FERMI-LAT (see \cite{paiano2017}).
\\

The determination of redshift of BLLs, especially in the distant Universe, is crucial and challenging.  The availability of large telescopes with state-of-the-art instrumentation improved the outcome of this hunt, as demostrated e.g. in \cite{shaw13, pita14, landoni2014, landonicfa, paiano2017b, paiano2017}. Nevertheless, only the advent of the Extremely Large Telescopes will greatly boost this line of research allowing to study both extremely beamed sources and most distance targets where faintest spectral lines from host galaxies could be revealed thanks to the giant jump in terms of SNR \citep{landoni2014}.

\newpage

\begin{table*} 
\caption{THE SAMPLE OF HIGH REDSHIFT SDSS BLLAC CANDIDATES}\label{tab:table1}
\begin{tabular}{lcccclccll}
\hline 
Object name     & RA          & $\delta$      & $l$ & $b$ & V      &  $E(B-V)$\footnote{Extinction values taken from NED}     &       Redshift\footnote{Redshift from Plotkin et al 2010. For SDSS J145059.98+520111.7 redshift is given by Liao et al 2015}       & Redshift Type  \\
                & (J2000)     & (J2000)    &            &               &               &               &  \\
\hline
SDSS J072659.52+373423.0	& 07:26:59.52 & +37:34:23.0  & 181.06 & 22.77 & 19.38 & 0.050 & 1.577 & T\\
SDSS J090818.95+214820.0	& 09:08:18.95 & +21:48:20.0  & 206.06 & 39.35 & 19.24 & 0.031 & 2.089 & T\\
SDSS J092902.42+194525.1	& 09:29:02.40 & +19:45:25.2  & 210.69 & 43.29 &19.25 & 0.050 & 1.774 & T\\
SDSS J094257.81-004705.2	& 09:42:57.82 & -00:47:05.3  & 236.68 & 36.82 & 19.00 & 0.045 & 1.360 & T	\\
SDSS J101115.63+010642.5	& 10:11:15.64 & +01:06:42.5  & 240.21 & 43.63 &19.40 & 0.041 & 1.479 & T \\
SDSS J120059.69+400913.1	& 12:00:59.68 & +40:09:13.1  & 158.44 & 73.30 & 19.90 & 0.200 & $\geq 3.370$ & LL \\
SDSS J121037.35+525341.9	& 12:10:37.34 & +52:53:41.8  & 136.61 & 63.13 & 18.80 & 0.020 & 0.917 & T    \\
SDSS J123132.37+013814.1	& 12:31:32.36 & +01:38:14.0  & 291.50 & 64.06 & 19.20 & 0.015 & $\geq 3.147$ & LL\\
SDSS J124030.93+344527.5	& 12:40:30.90 & +34:45:27.5  & 139.30 & 82.02 &19.35 & 0.010 & 1.642 & T\\
SDSS J124700.72+442318.8	& 12:47:00.72 & +44:23:18.8  & 125.59 &  72.71 & 18.90 & 0.020 & 1.812 & T\\
SDSS J130925.52+430505.5	& 13:09:25.52 & +43:05:05.6  & 111.20 & 73.63 & 17.30 & 0.019 & 1.154 & T 	\\
SDSS J132802.09+112913.6	& 13:28:02.09 & +11:29:13.7  & 333.48 & 72.15 &18.53 & 0.025 & 1.827 & T\\
SDSS J144050.14+333350.2	& 14:40:50.14 & +33:33:50.1  & 54.91 & 65.61 & 18.40 & 0.100 & 1.774 & T\\
SDSS J145059.98+520111.7\footnote{Published in Paiano et. al. 2017}	& 14:50:59.98 &+52:01:11.7 & 89.13 & 56.53 & 18.90 & 0.060 & $\geq 2.471$& LL \\
SDSS J152422.56+374034.1	& 15:24:22.56 & +37:40:34.2  & 60.934 & 56.30& 19.80 & 0.034 & 1.219 & T \\
SDSS J170108.89+395443.0	& 17:01:08.89 & +39:54:43.0  & 63.96 & 37.48 & 19.30 & 0.024 & 1.889 & T \\

\hline

\end{tabular}
\footnotesize {\texttt{Col.1}: Name of the target; \texttt{Col.2}: Right Ascension; \texttt{Col.3}: Declination; \texttt{Col.4 }: Galactic coordinate l ;\texttt{Col.5 }: Galactic coordinate b ; \texttt{Col.6}: Magnitude from catalog; \texttt{Col.7: } Reddening (from NED); \texttt{Col.8:} Redshift ; \texttt{Col.9} Type of redshift: T - tentative; LL - Lower limit from intervening systems \\}

\end{table*} 

\begin{table*}
\caption{LOG OBSERVATIONS OF HIGH-Z SDSS SOURCES OBTAINED AT GTC}\label{tab:table2}
\centering
\begin{tabular}{llcllllc}
\hline
  \multicolumn{4}{c}{Grism B}    & \multicolumn{4}{c}{Grism R}   \\
\hline
Object          & t$_{Exp}$ (s)  &       Date            & Seeing ($^{\prime\prime}$) & t$_{Exp}$ (s) &         Date         & Seeing ($^{\prime\prime}$) & r  \\
\hline
SDSS J072659.52+373423.0	& 1500  & 29-01-2015 & 2.2  & 2400 & 29-01-2015 & 2.1 & 19.80 \\
SDSS J090818.95+214820.0	& 1800  & 30-01-2015 & 2.1  & 2500 & 30-01-2015 & 1.8 & 19.15 \\
SDSS J092902.42+194525.1	& 1500  & 14-03-2015 & 2.1  & 2400 & 14-03-2015 & 1.9 & 19.60 \\
SDSS J094257.81-004705.2	& 1250  & 10-02-2015 & 1.7  & 1800 & 10-02-2015 & 1.9 & 19.00 \\
SDSS J101115.63+010642.5	& 1800  & 11-02-2015 & 1.5  & 2500 & 11-02-2015 & 1.4 & 18.90 \\
SDSS J120059.69+400913.1	& 7200  & 24-02-2015 & 1.8  & 7200 & 28-02-2015 & 2.1 & 19.50 \\
SDSS J121037.35+525341.9	& 900   & 12-02-2015 & 2.8  & 1800 & 12-02-2015 & 2.3 & 19.20 \\
SDSS J123132.37+013814.1	& 1500  & 30-01-2015 & 2.0  & 2100 & 30-01-2015 & 1.8 & 18.80 \\
SDSS J124030.93+344527.5	& 1800  & 12-02-2015 & 1.7  & 2400 & 12-02-2015 & 1.9 & 19.00 \\
SDSS J124700.72+442318.8	& 1300  & 12-02-2015 & 2.5  & 1500 & 12-02-2015 & 2.4 & 19.10 \\
SDSS J130925.52+430505.5	& 300   & 04-02-2015 & 1.0  & 700  & 04-02-2015 & 0.9 & 17.00 \\
SDSS J132802.09+112913.6	& 600   & 04-02-2015 & 1.1  & 1500 & 04-02-2015 & 0.9 & 18.50 \\
SDSS J144050.14+333350.2	& 900   & 04-02-2015 & 0.9  & 900  & 04-02-2015 & 0.8 & 18.60 \\
SDSS J145059.98+520111.7	& 1500  & 14-03-2015 & 2.0  & 1800 & 14-03-2015 & 1.9 & 18.90 \\
SDSS J152422.56+374034.1	& 2400  & 25-02-2015 & 1.2  & 3000 & 14-03-2015 & 1.8 & 19.40 \\
SDSS J170108.89+395443.0	& 2400  & 14-03-2015 & 2.1  & 2400 & 14-03-2015 & 2.0 & 19.20 \\

\hline

\end{tabular}

\footnotesize {\texttt{Col.1}: Name of the target; \texttt{Col.2}: Exposure time for R1000B grism (sec); \texttt{Col.3}: Date of observation of grism R1000B; \texttt{Col.4 }: Seeing during acquisition of R1000B spectrum ;\texttt{Col.5 }: Exposure time for R1000R grism;  \texttt{Col.6}:Date of observation of grism R1000R; \texttt{Col.7: } Seeing during acquisition of R1000R spectrum; \texttt{Col.8:} Magnitude measured from acquisition image in r band \\} 
\end{table*}

\begin{table*}
\caption{PROPERTIES OF THE OPTICAL SPECTRA OF HIGH Z SOURCES}\label{tab:table3}
\centering
\begin{tabular}{ l ccll}
\hline
OBJECT                       &   SNR GTC         &   EW$_{min}$   &  z   & Class                                 \\
\hline
SDSS J072659.52+373423.0 & 20 & 1.00$-$ 1.50& 0.791$^{(g)}$ & B\\
SDSS J090818.95+214820.0  & 30 & 0.65$-$ 1.20 & 2.088 & B\\ 
SDSS J092902.42+194525.1   & 25 & 0.70$-$ 1.10 & $\geq 0.35$$^{(ll)}$ & B\\ 
SDSS J094257.81-004705.2  & 35 & 0.60$-$ 0.90 & 1.363 & B\\ 
SDSS J101115.63+010642.5  & 50 & 0.30$-$ 0.80 & 0.857 & Q \\ 
SDSS J120059.69+400913.1 &  25 & 0.70$-$ 1.30 & $\geq 3.367$$^{(i)}$ & B\\ 
SDSS J121037.35+525341.9 &  40 & 0.30$-$ 1.10 &  0.916 & Q \\ 
SDSS J123132.37+013814.1 &  30 & 0.40$-$ 1.00 & $\geq 3.140$$^{(i)}$& B \\
SDSS J124030.93+344527.5 &  50 & 0.37$-$ 0.80 & 1.636 & B\\  
SDSS J124700.72+442318.8 &  60 & 0.20$-$ 0.90 & 0.569 ? & B \\ 
SDSS J130925.52+430505.5 &  100 & 0.55$-$ 0.90 & 0.693 & B\\ 
SDSS J132802.09+112913.6 &  30 & 0.55$-$ 1.00 & 0.580 & Q \\ 
SDSS J144050.14+333350.2 &  30 & 0.50$-$ 0.90 & 1.747 & Q \\ 
SDSS J145059.98+520111.7 &  70 & 0.50$-$ 0.90 & $\geq 2.470$$^{(i)}$ & B\\ 
SDSS J152422.56+374034.1 &  40 & 0.60$-$ 1.00 & 1.219 & Q\\
SDSS J170108.89+395443.0 &  25 & 0.60$-$ 1.00 & 1.895& Q \\ 

\hline
\end{tabular}
\tablenotetext{}{
\raggedright
\footnotesize \texttt{Col.1}: Name of the target; \texttt{Col.2}: Average GTC S/N of the spectrum; \texttt{Col.3}: Range of the minimum equivalenth width (EW$_{min}$) derived from different regions of the spectrum (see \cite{paiano2017b}), \texttt{Col.4}: Proposed redshift from this work. The superscript letters indicate \textit{g} = host galaxy absorption, \textit{i}= intervening absorption, \textit{ll} = lower limit  of the redshift by assuming a BL Lac host galaxy with $M_{R}$ = -22.9; \texttt{Col.5}: Class of the object (B: BLL, Q: FSRQ)}
\end{table*}

\begin{table*}
\caption{MEASUREMENTS OF SPECTRAL LINES}\label{tab:line}
\centering
\begin{tabular}{llcllcc}
\hline
OBJECT            &  $\lambda_{obs}$    &    EW (observed)    &     Line ID    &   z$_{line}$   & Luminosity & Type \\
                  &  $\textrm{\AA}$             &     $\textrm{\AA}$           &                &              & log (L$_{\odot}$)                    &               \\
\hline
SDSS J072659.52+373423.0 & 7043.80 & 2.60 & Ca II (3934) & 0.791 &  & G \\

	& 7107.40 & 1.80 & Ca II (3968) & 0.791&  & G \\	
	\hline
 SDSS J090818.95$+$214820.0                                       &   4789.20     &   20.10       &      C IV  (1550)      &  2.088 & 43.45 & E\\ 
        &   8650.40    &   19.30         &      Mg II  (2800)      &  2.089 &43.43& E \\ 
         	\hline
SDSS  J094257.81$-$004705.2      & 4509.20 & 14.90 & C III] (1909) & 1.362 & 42.90 & E\\
					&   5086.39      &  2.30        &      Mg II (2796)       &   0.819 && A\\
                                        &   5100.18   &  2.10         &      Mg II (2803)       &   0.819& &A \\
                                        &   6608.92     &  10.30         &    Mg II (2800)     &  1.363  & 42.75& E\\
                                        	\hline
 SDSS J101115.63+010642.5 & 6310.00 & 3.50 & [Ne V] (3346) & 0.885$^{*}$ & & E\\
 					      & 6929.00 & 3.00 & [O II] (3727) & 0.858$^{*}$ & & E\\
					      & 9028.00 & 2.70 & H$\beta$ (4861) & 0.857$^{*}$ & & E \\
					      & 9209.00 & 30.00 & [O III] (4959) & 0.857$^{*}$ & & E\\
					      & 9298.00 & 50.00 & [O III] (5007) & 0.857$^{*}$ &  & E\\
				\hline		      
SDSS J120059.69+400913.1 & 5310.20 & 56.10 & Ly-$\alpha$ (1216) & 3.367 & & A \\
					      & 5991.40 & 2.50 & Mg II (2796) & 1.142&& A \\
					      & 6003.20 & 2.00 & Mg II (2803) & 1.142&& A \\
					      & 5919.62 & 2.50 & Fe II (2382) & 1.484&& A \\
					      & 6459.50 & 2.00 & Fe II (2600) & 1.484&& A \\
					      & 6946.50 & 4.50 & Mg II (2796) & 1.484&& A \\
					      & 6964.10 & 3.00 & Mg II (2803) & 1.484&& A \\
	\hline
SDSS J121037.35$+$525341.9      &   5360.20     &  29.10         &    Mg II (2800)     &  0.917 & 43.60 & E\\
       &   7142.40    & 3.70       &    [O II] (3727)   & 0.916          & 42.60 & E        \\ 
       	\hline
       
 SDSS J123132.37+013814.1  & 5434.00 & 49.60 & Ly-$\alpha$ (1216) & 3.140 && A\\ 
 						& 5527.20 & 2.00 & Mg II (2796) & 0.977&& A\\
 					       & 5541.50 & 1.50 & Mg II (2803) & 0.977&& A \\
					       & 6154.00 & 3.50 & Mg II (2796) & 1.200&& A \\
					       & 6167.10 & 1.50 & Mg II (2803) & 1.200&& A \\
					       & 7812.20 & 1.50 & Fe II (2600) & 2.004&& A \\
					       & 8401.30 & 5.00 & Mg II (2796) & 2.004 && A\\
					       & 8422.20 & 3.50 & Mg II (2803) & 2.004 && A\\
       	\hline
 SDSS J124030.93$+$344527.5  & 5741.10     &  1.70         &     ?     &  ? && A\\
 						& 5815.50     &  1.50         &     ?     &  ? && A\\
						& 5835.80     &  2.00         &     Fe II  (2382)   &  1.449 && A\\
						& 6335.20     &  2.20         &     ?     &  ? && A\\
						& 6368.60     &  5.00         &     Fe II (2600)     &  1.449&& A \\
						& 6848.90     &  5.00         &     Mg II (2796)     &  1.449&& A \\
						& 6866.30     &  7.00         &     Mg II (2803)     &  1.449 && A\\
 						&   7390.10     &  12.30         &    Mg II (2800)      &  1.636  & 43.00 & E\\
 	\hline
 SDSS J124700.72$+$442318.8 &   7858.20     &  2.30         &    [O III] (5007) ?     &  0.569 ? & 41.00 & E \\
 	\hline
   SDSS J130925.52$+$430505.5 &  4724.55 & 0.40 & Mg II (2796) & 0.690&& A \\
     & 4736.70 & 0.20 & Mg II (2803) & 0.690&& A \\
  & 6309.35$^{*}$ & 1.70 & [O II] (3727) & 0.693 & 42.00 & E\\
  &   8474.30    & 0.50       &    [O III] (5007)   & 0.693 & 41.40 & E                  \\ 
	\hline

\hline
\end{tabular}
\tablenotetext{}{
\raggedright
\footnotesize \texttt{Col.1}: Name of the target; \texttt{Col.2}: Barycenter of the detected line ; \texttt{Col.3}: Measured equivalent width; \texttt{Col.4}: Line identification; \texttt{Col.5}: Spectroscopic redshift. (the $*$ indicates that value is determined from barycenter of detected lines contaminated by telluric absorptions); \texttt{Col.6}: Line luminosity (only for emission features). \texttt{Col.7}:Feature type: G absorption from host galaxy, E emission line while A denotes absorption lines.}
\end{table*}
\setcounter{figure}{3}
\setcounter{table}{3}
\begin{table*}
\caption{MEASUREMENTS OF SPECTRAL LINES (continued)}\label{tab:line}
\centering
\begin{tabular}{llcllcc}
\hline
OBJECT            &  $\lambda_{obs}$    &    EW (observed)    &     Line ID    &   z$_{line}$   & Luminosity & Type \\
                  &  $\textrm{\AA}$             &     $\textrm{\AA}$           &                &       & log (L$_{\odot}$)                                          \\
\hline

SDSSJ132802.09$+$112913.6 & 4409.90 & 86.70 & Mg II (2800) & 0.577  & 42.90& E\\
  &   5410.40    & 3.50       &    [Ne V] (3426)   & 0.578 & 41.50& E\\
  &   7915.10    & 6.50       &    [O III] (5007)   & 0.580 & 41.80& E \\
  	\hline
  SDSS J144050.14$+$333350.2 & 4254.90 & 15.30 & C IV (1550) & 1.745 & 43.35 & E\\
  & 4414.15 & 1.35 & Mg II (2796) & 0.578 && A \\
  & 4426.24 & 1.20 & Mg II (2803) & 0.579&& A \\
  & 5245.80 & 30.35 & C III] (1909) &  1.747 & 43.70 &E \\
    
  & 6063.04 & 2.20 & Mg II (2803) & 1.168 && A \\
  & 6078.50 & 1.30 & Mg II (2796) & 1.168  && A\\
  & $\sim$ 7700 & 18.00 & Mg II (2800) & $\sim$1.74$^{*}$ &  & E\\
  	\hline
  SDSS J152422.56$+$374034.1 & 4235.40 & 5.30 &  C III] (1909)& 1.218 & 42.10 & E\\
    & 6204.50 & 64.30 &  Mg II (2800)& 1.219 & 42.90 & E\\
	\hline
 SDSS J170108.89+395443.0   & 4490.50 & 30.00 &C IV (1550) & 1.895 & 43.40 &E \\
   & 8120.50 & 11.10 &  Mg II (2800) & 1.896 & 42.90 & E\\ 

\hline
\end{tabular}
\tablenotetext{}{
\raggedright
\footnotesize \texttt{Col.1}: Name of the target; \texttt{Col.2}: Barycenter of the detected line ; \texttt{Col.3}: Measured equivalent width; \texttt{Col.4}: Line identification; \texttt{Col.5}: Spectroscopic redshift. (the $*$ indicates that value is determined from barycenter of detected lines contaminated by telluric absorptions); \texttt{Col.6}: Line luminosity (only for emission features). \texttt{Col.7}:Feature type: G absorption from host galaxy, E emission line while A denotes absorption lines.}
\end{table*}

\bibliographystyle{aasjournal}
\bibliography{SDSS_GTC_biblio}

\newpage

\setcounter{figure}{0}
\begin{figure*}
	\centering
	\includegraphics[width=15cm, angle=270]{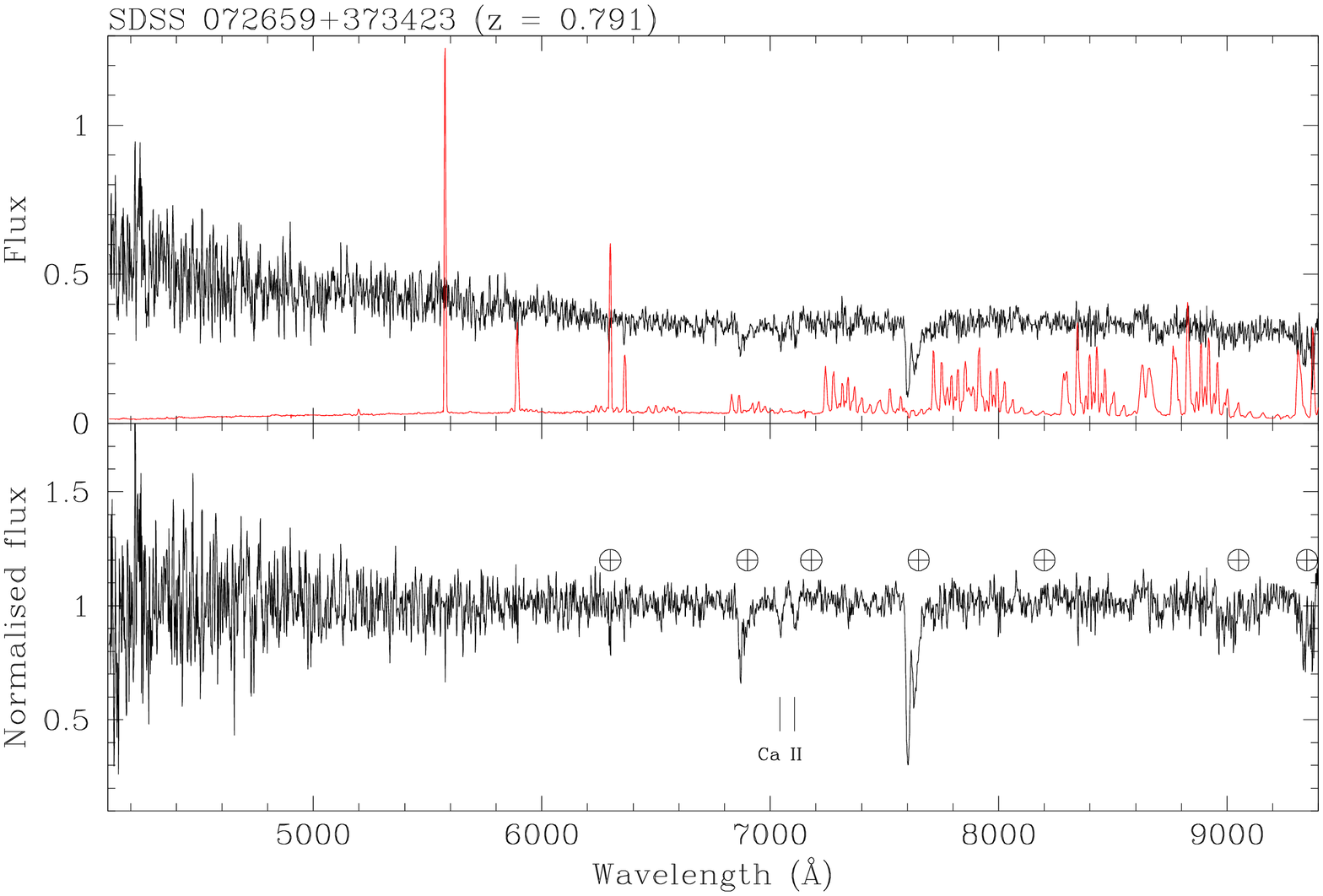}
    \caption{\textit{Top panel}: Black solid line is the flux calibrated and deredded spectra while the red line is the sky spectrum in arbitrary units reported for sake of comparison. \textit{Bottom panel}: Normalized spectra. The main telluric bands are indicated by $\oplus$, the absorption features from interstellar medium of our galaxies are labelled as IS (Inter-Stellar). Data available at URL: http://www.oapd.inaf.it/zbllac/}
    \label{fig:example_pair}
\end{figure*}

\begin{figure*}
	\centering
	\includegraphics[width=15cm, angle=270]{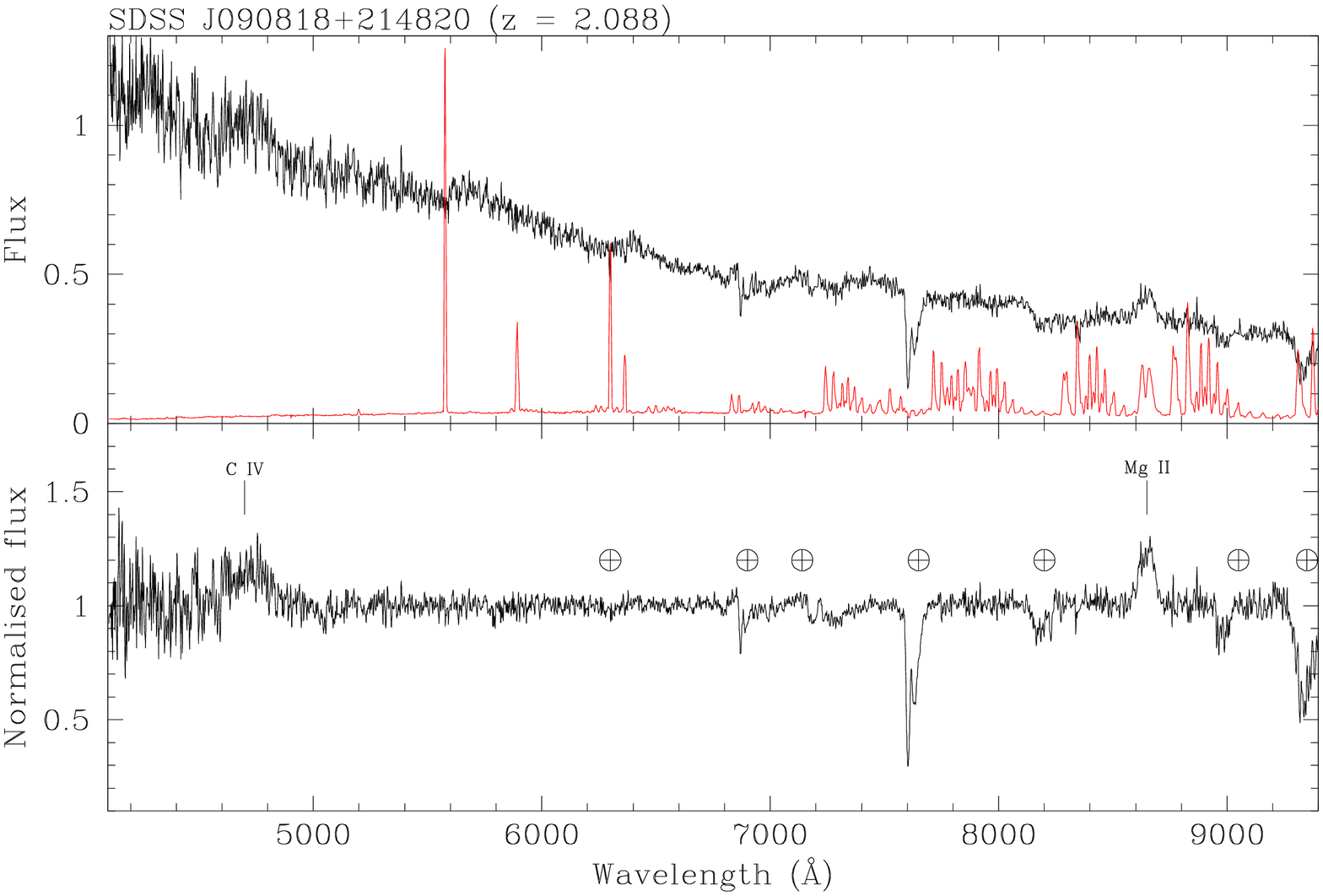}
    \label{fig:example_pair}
\end{figure*}

\begin{figure*}
	\centering
	\includegraphics[width=15cm, angle=270]{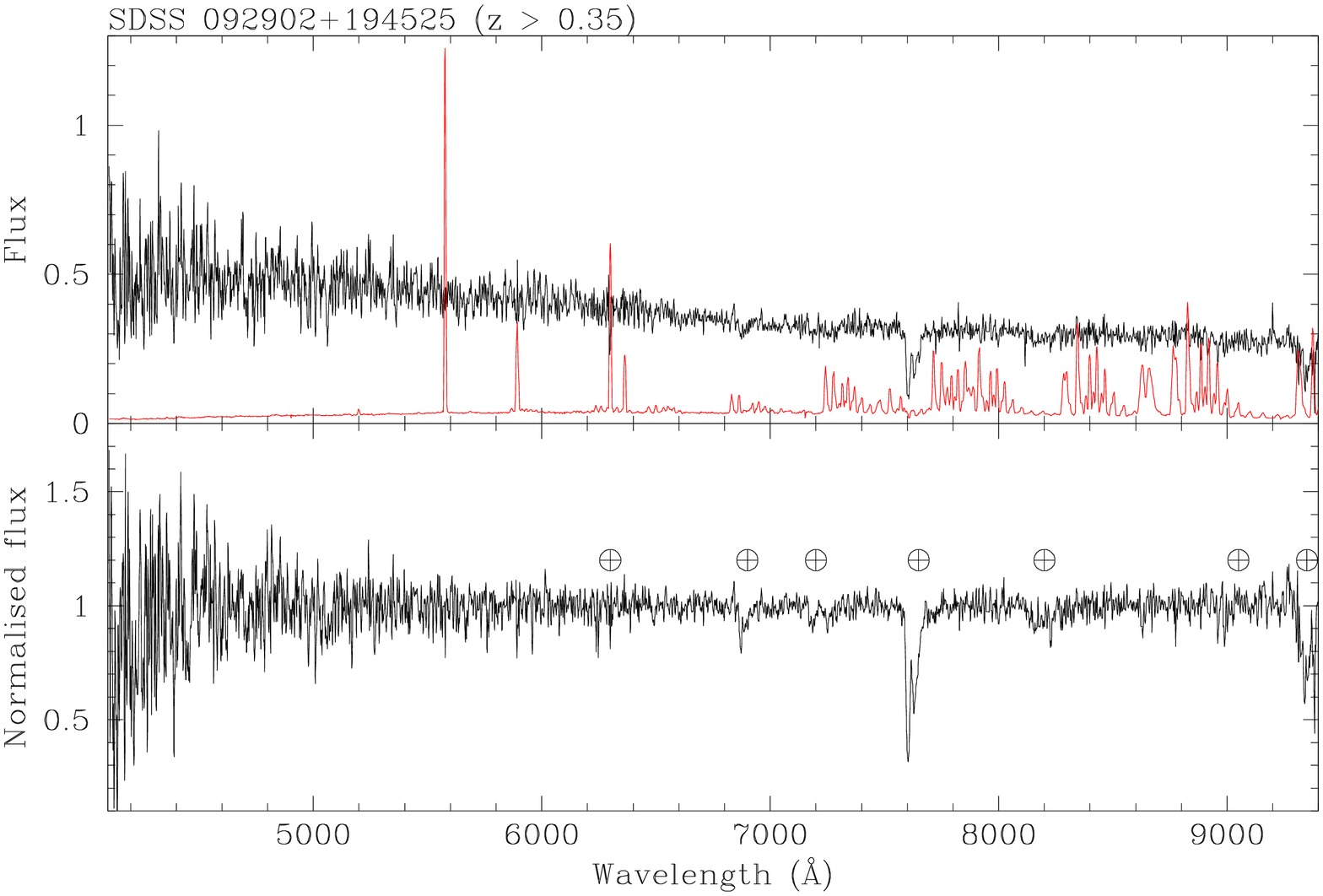}
    \label{fig:example_pair}
\end{figure*}

\begin{figure*}
	\centering
	\includegraphics[width=15cm, angle=270]{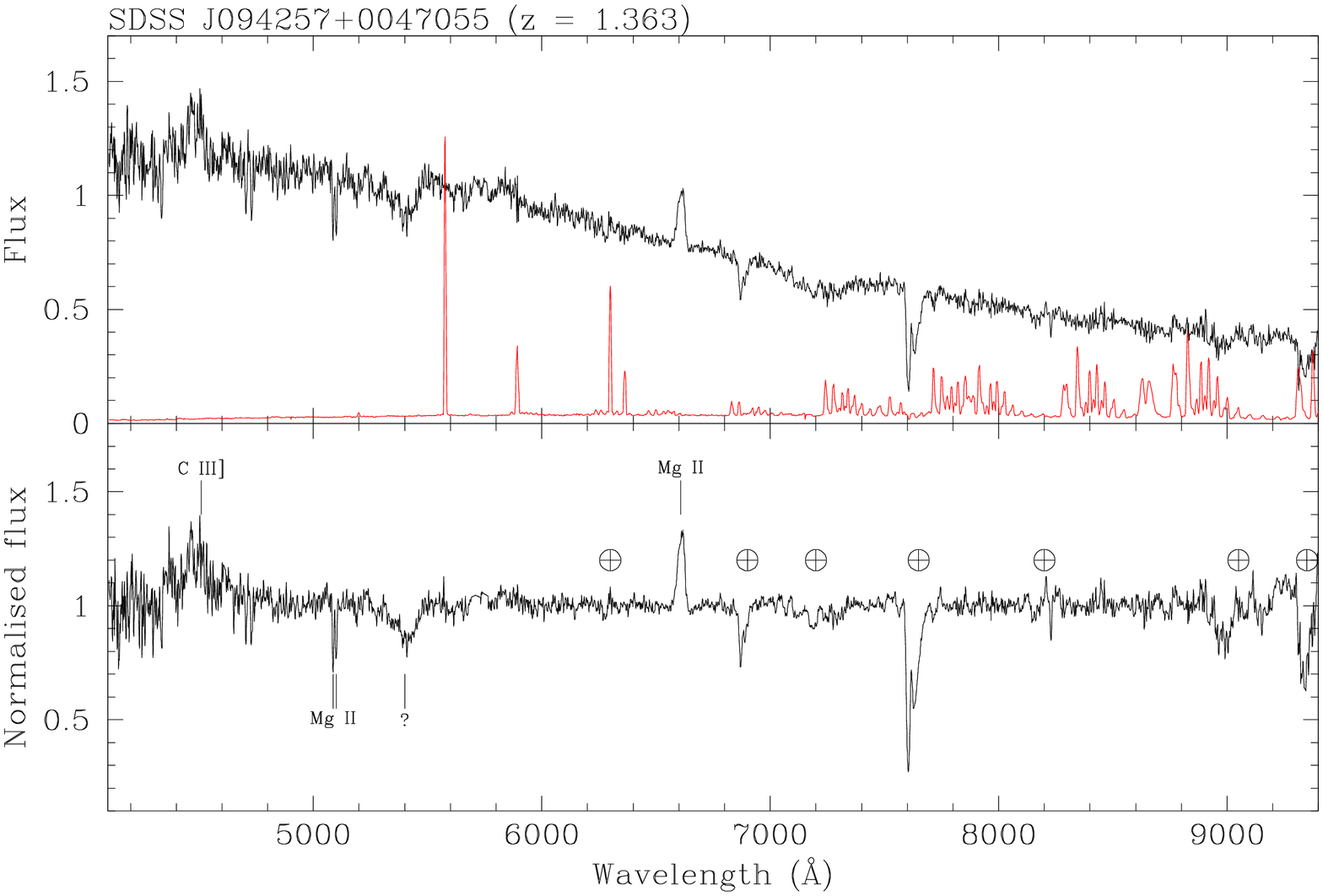}
    \label{fig:example_pair}
\end{figure*}

\begin{figure*}
	\centering
	\includegraphics[width=15cm, angle=270]{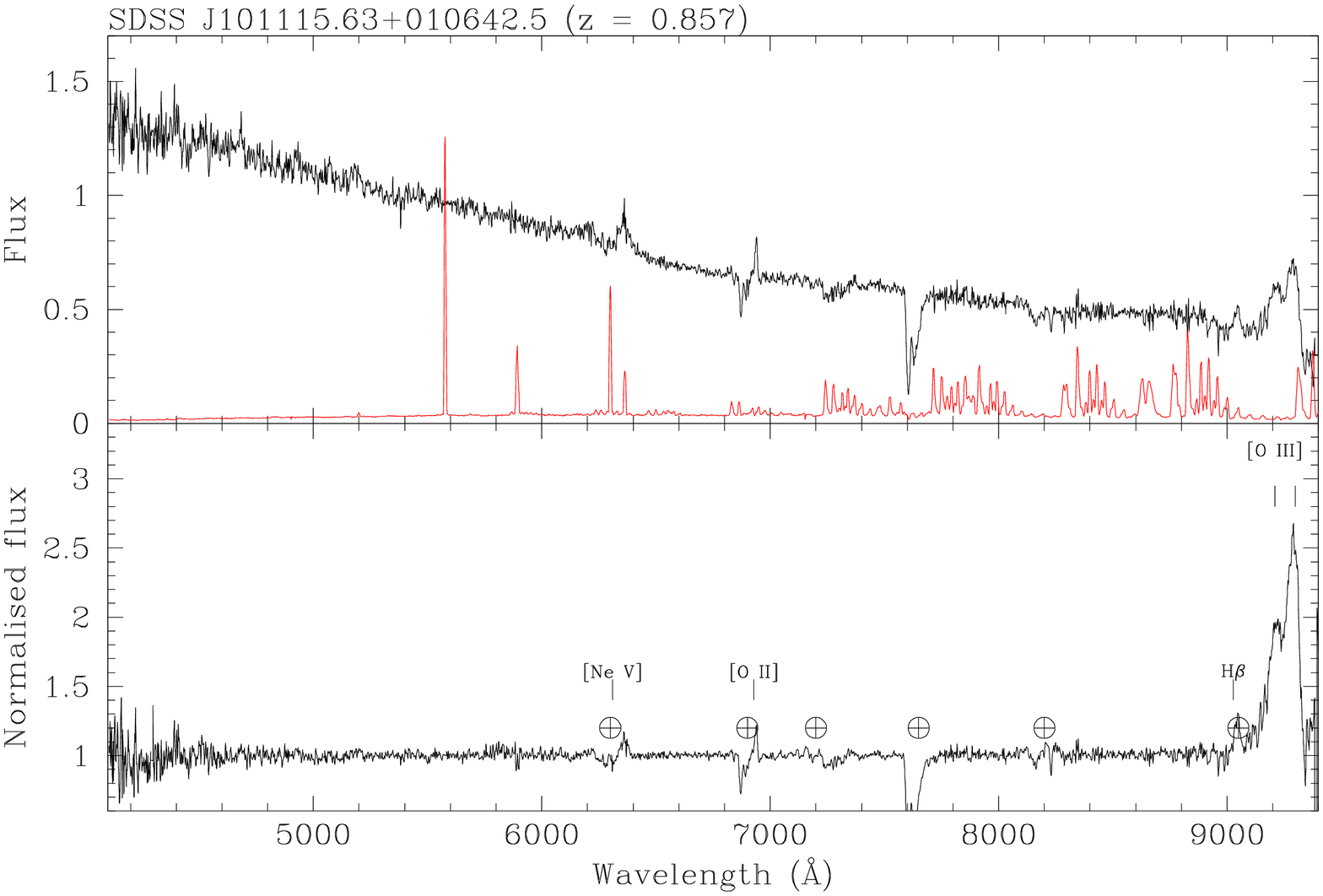}
    \label{fig:example_pair}
\end{figure*}

\begin{figure*}
	\centering
	\includegraphics[width=15cm, angle=270]{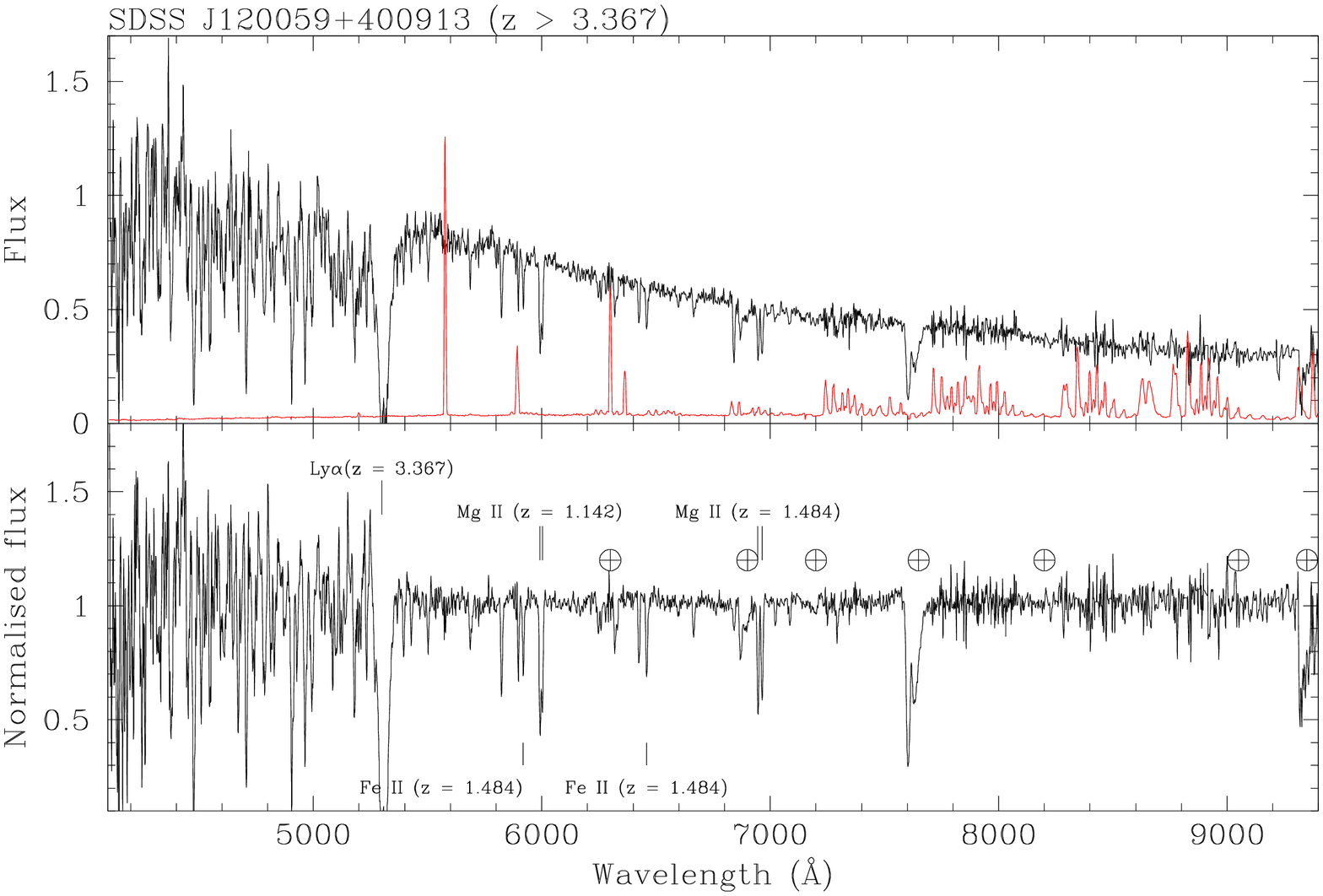}
    \label{fig:example_pair}
\end{figure*}

\clearpage
\begin{figure*}
	\centering
	\includegraphics[width=15cm, angle=270]{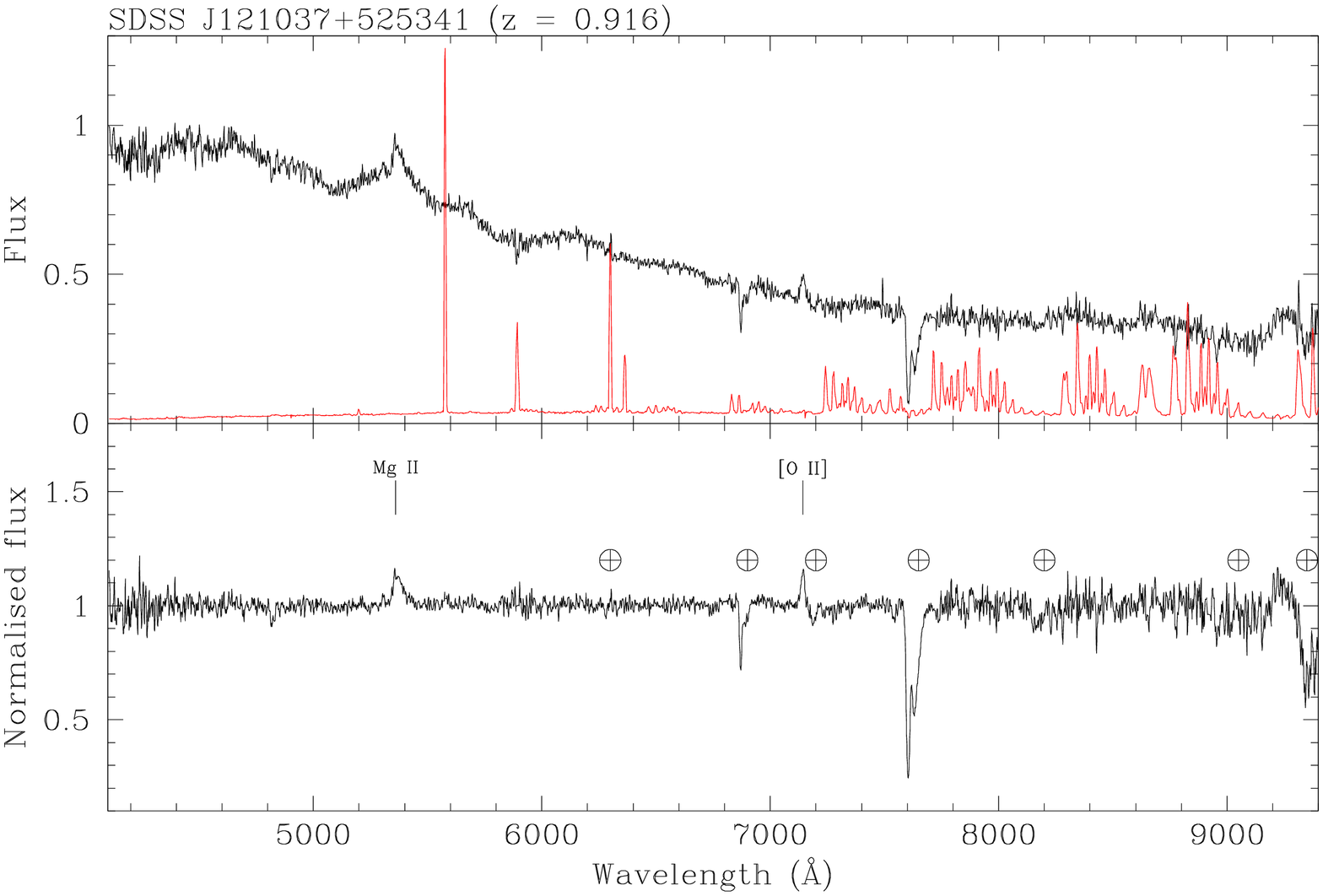}
    \label{fig:example_pair}
\end{figure*}

\begin{figure*}
	\centering
	\includegraphics[width=15cm, angle=270]{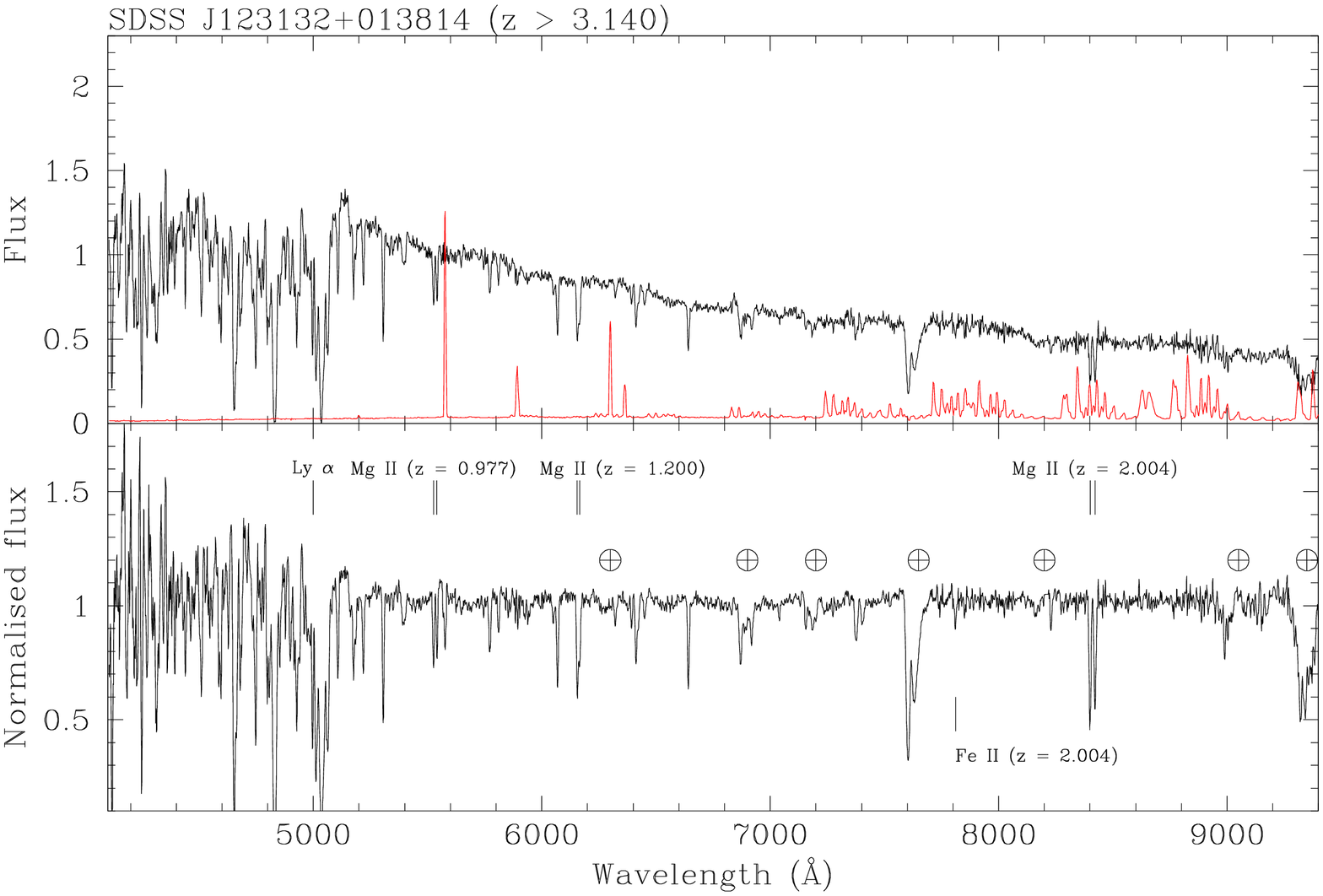}
    \label{fig:example_pair}
\end{figure*}

\begin{figure*}
	\centering
	\includegraphics[width=15cm, angle=270]{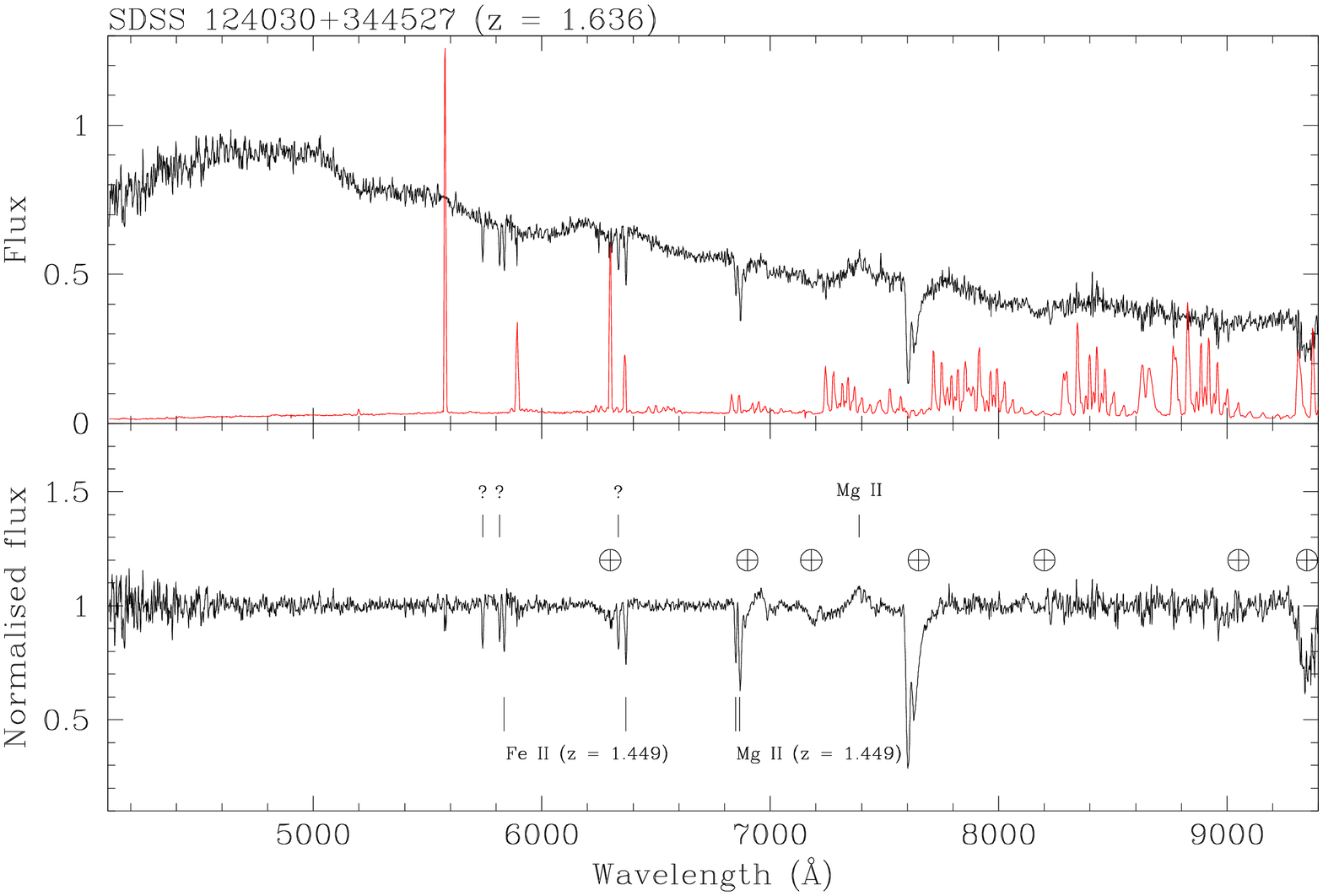}
    \label{fig:example_pair}
\end{figure*}

\begin{figure*}
	\centering
	\includegraphics[width=15cm, angle=270]{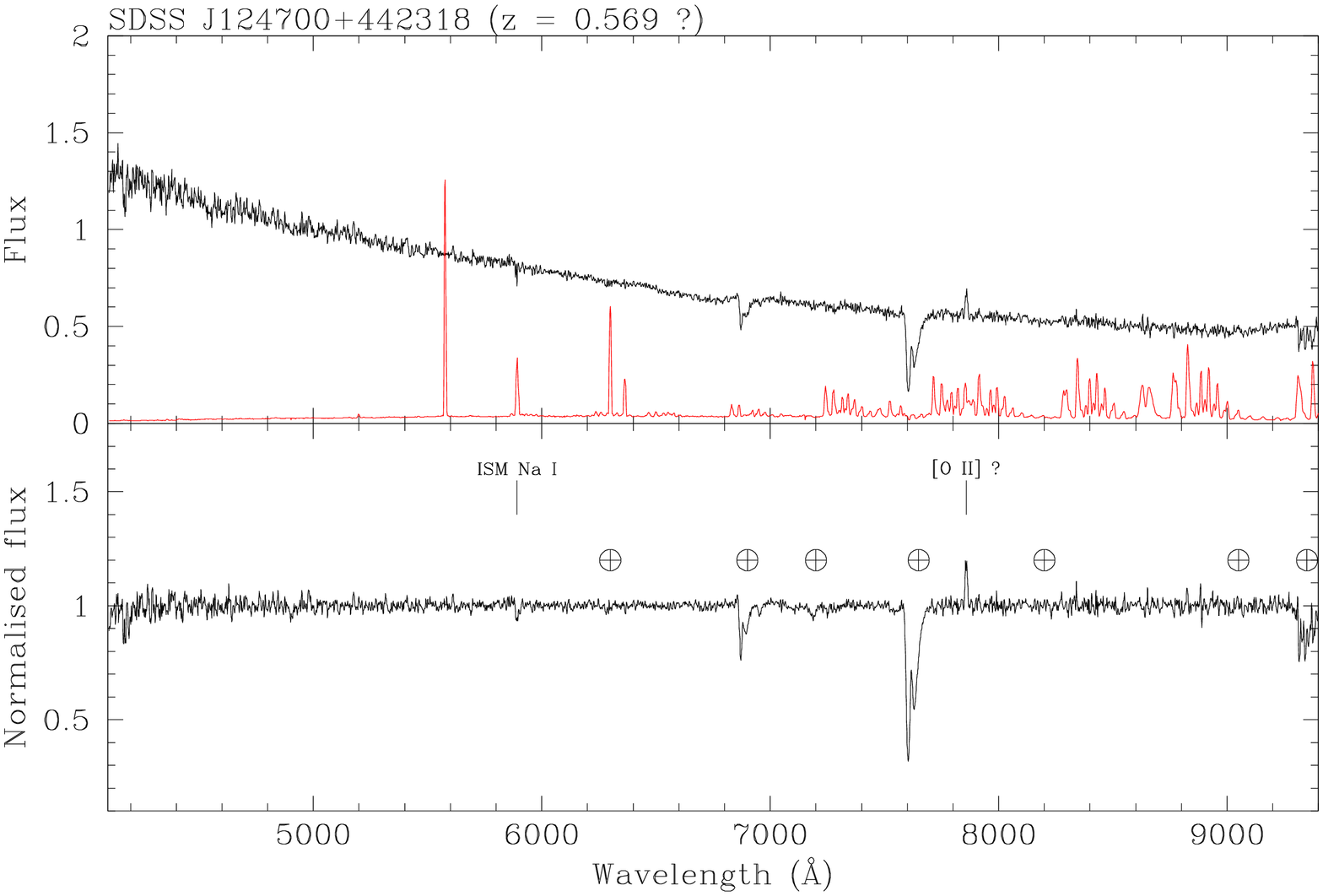}
    \label{fig:example_pair}
\end{figure*}

\begin{figure*}
	\centering
	\includegraphics[width=15cm, angle=270]{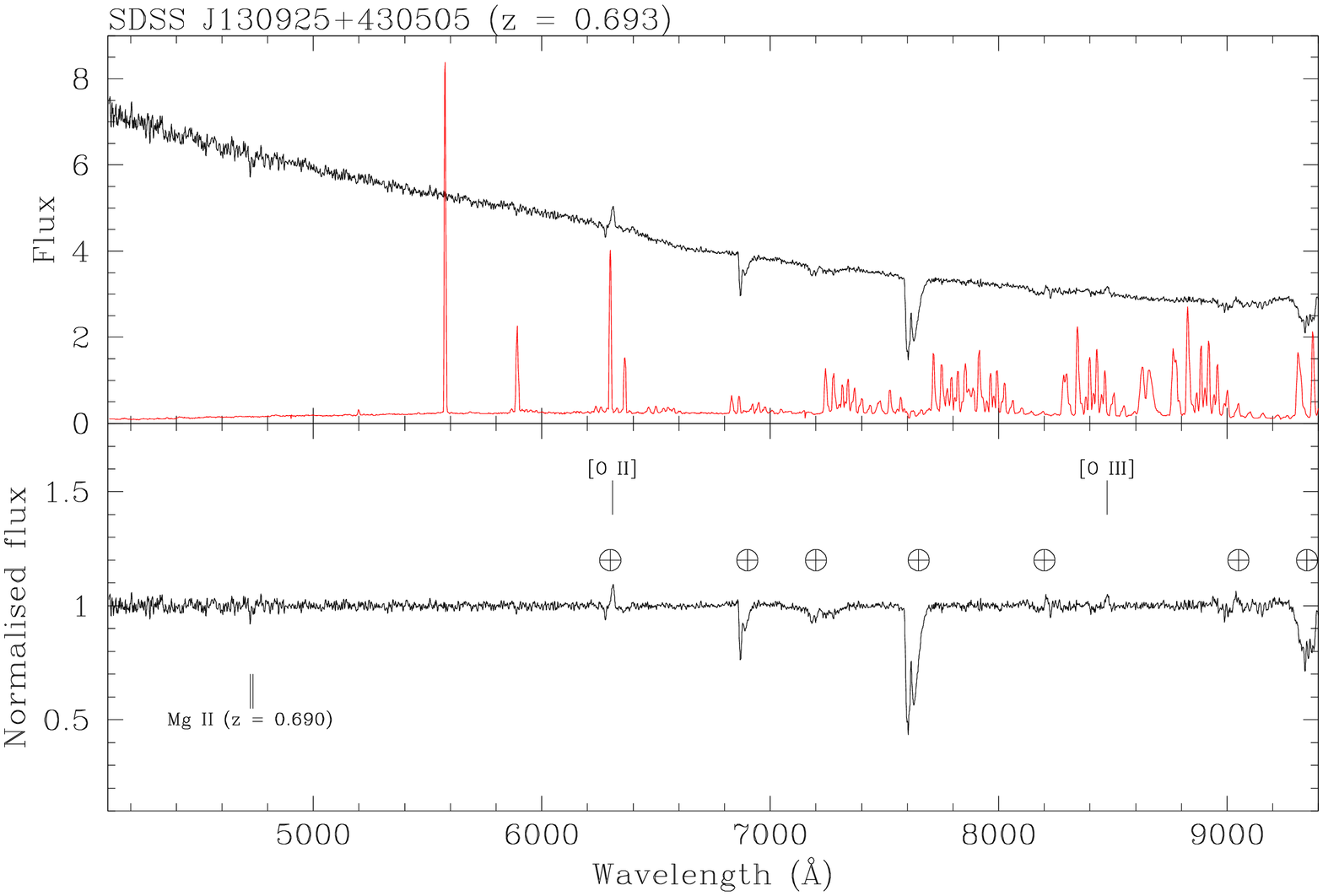}
    \label{fig:example_pair}
\end{figure*}
\clearpage

\begin{figure*}
	\centering
	\includegraphics[width=15cm, angle=270]{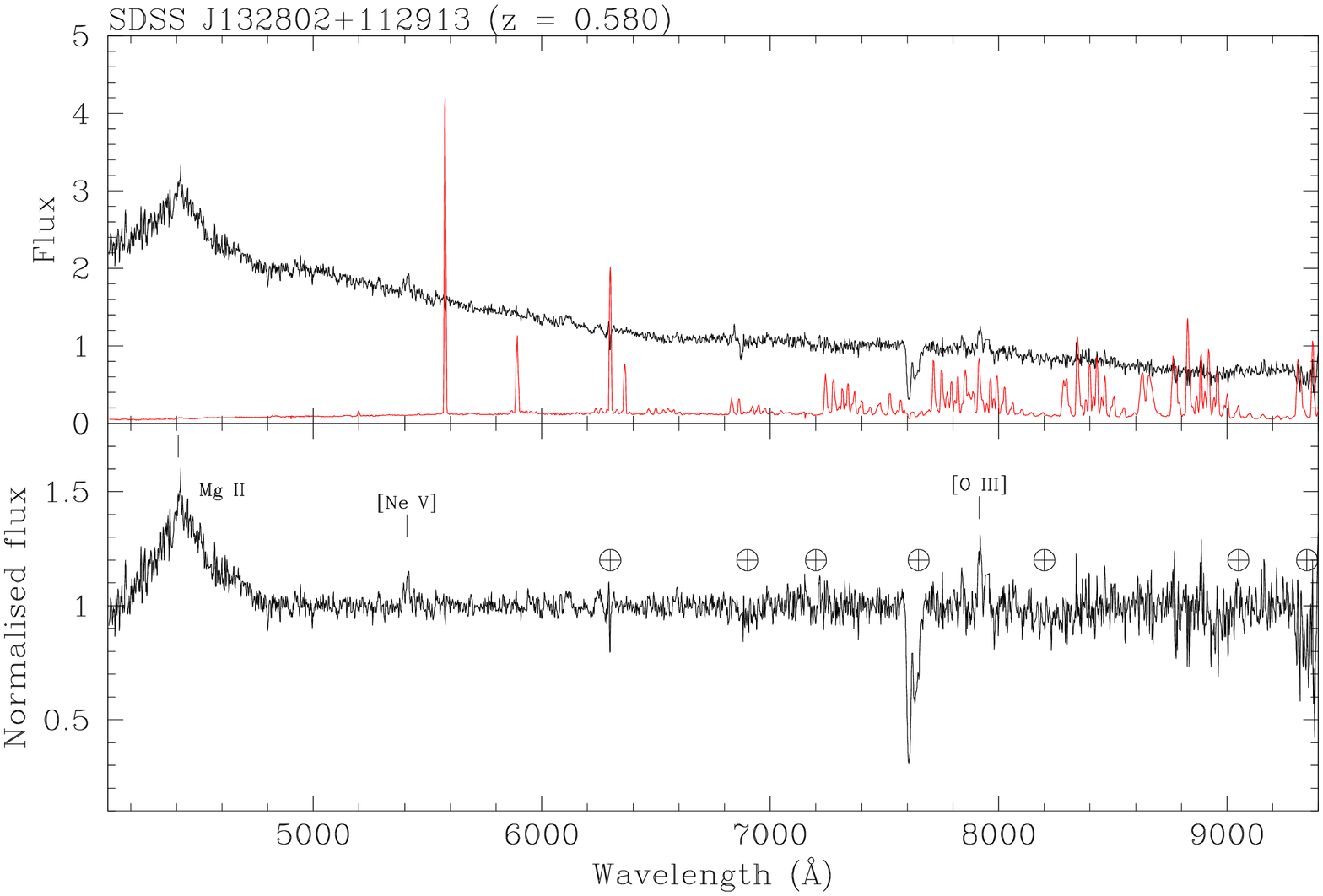}
    \label{fig:example_pair}
\end{figure*}

\begin{figure*}
	\centering
	\includegraphics[width=15cm, angle=270]{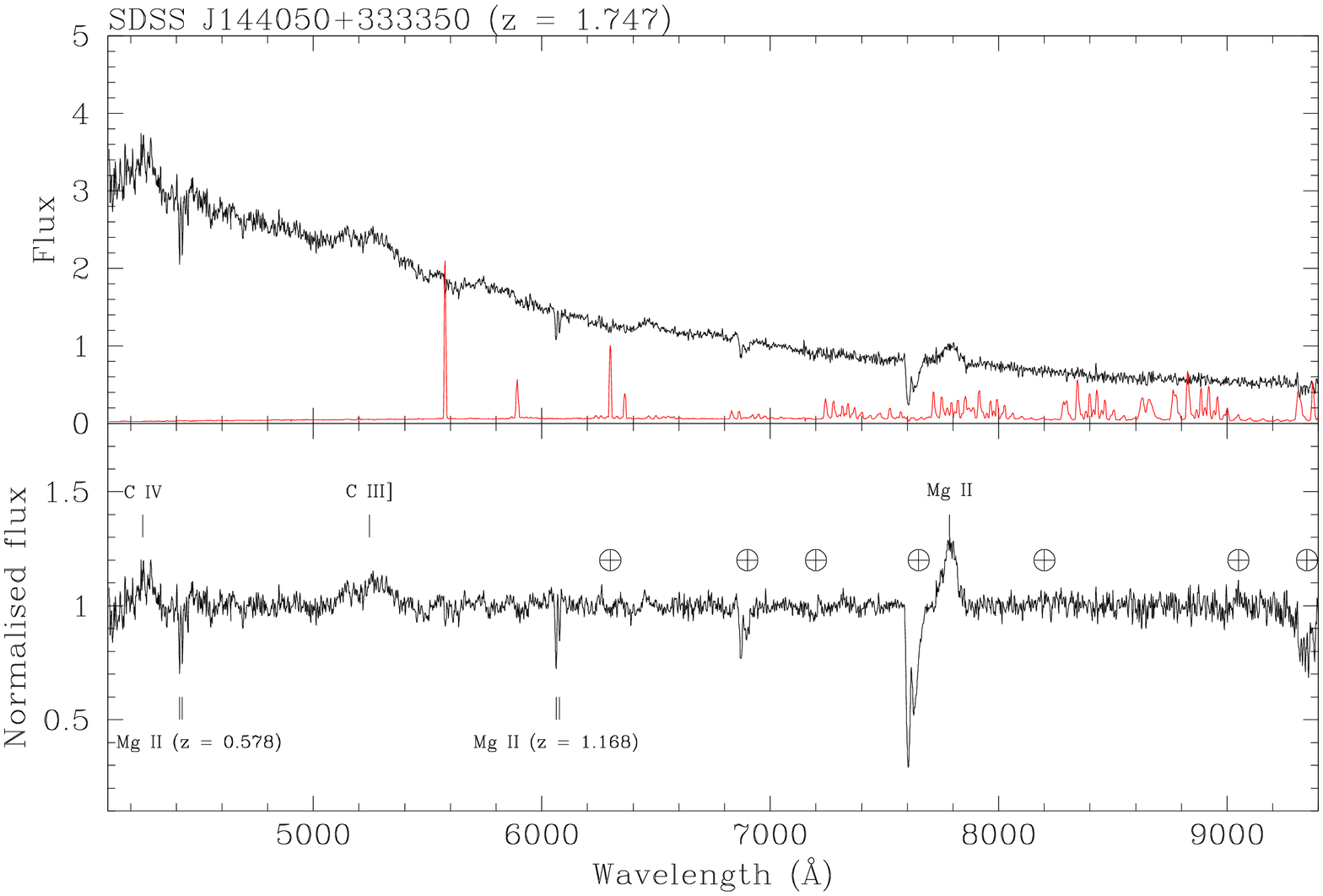}
    \label{fig:example_pair}
\end{figure*}

\begin{figure*}
	\centering
	\includegraphics[width=15cm, angle=270]{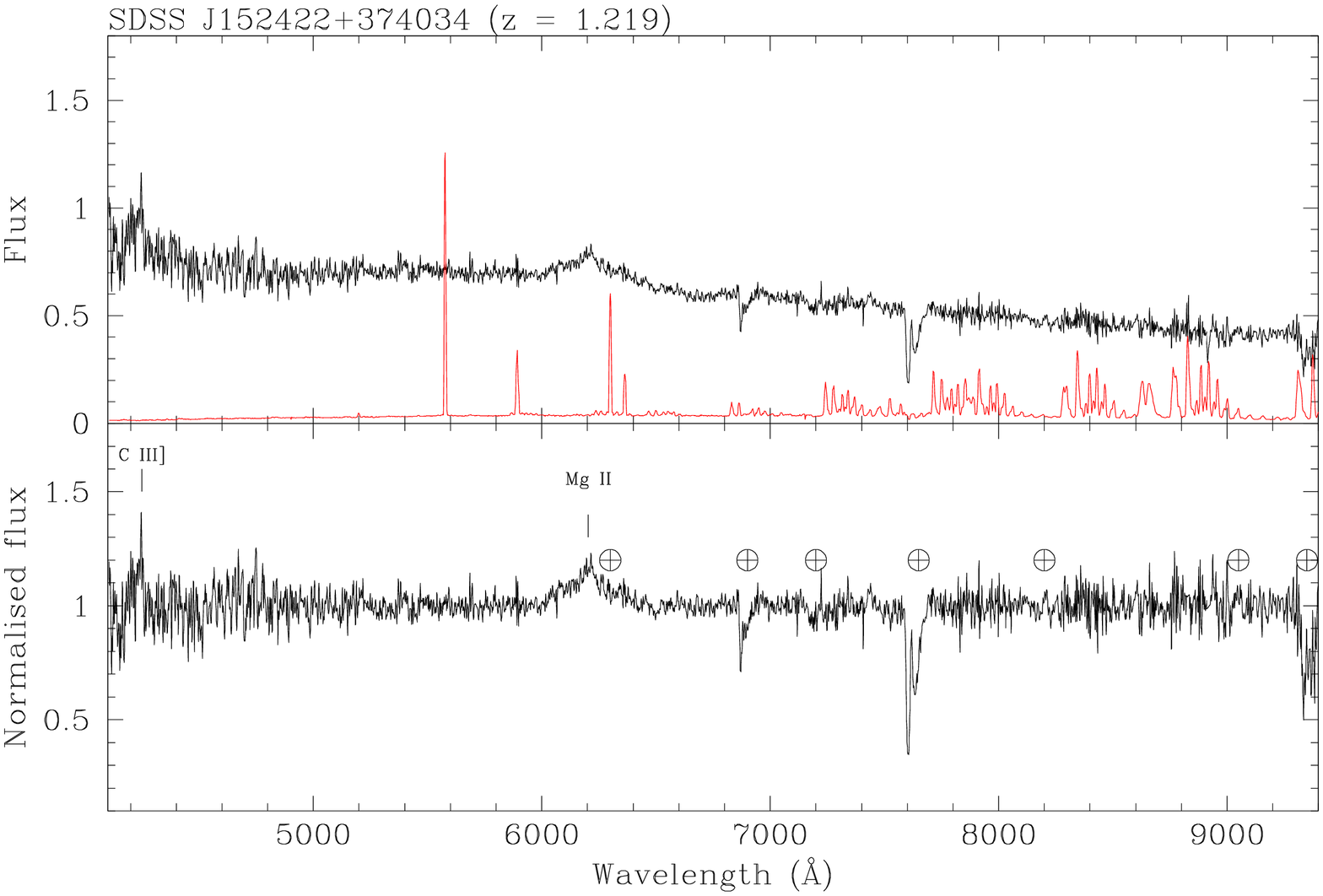}
    \label{fig:example_pair}
\end{figure*}

\begin{figure*}
	\centering
	\includegraphics[width=15cm, angle=270]{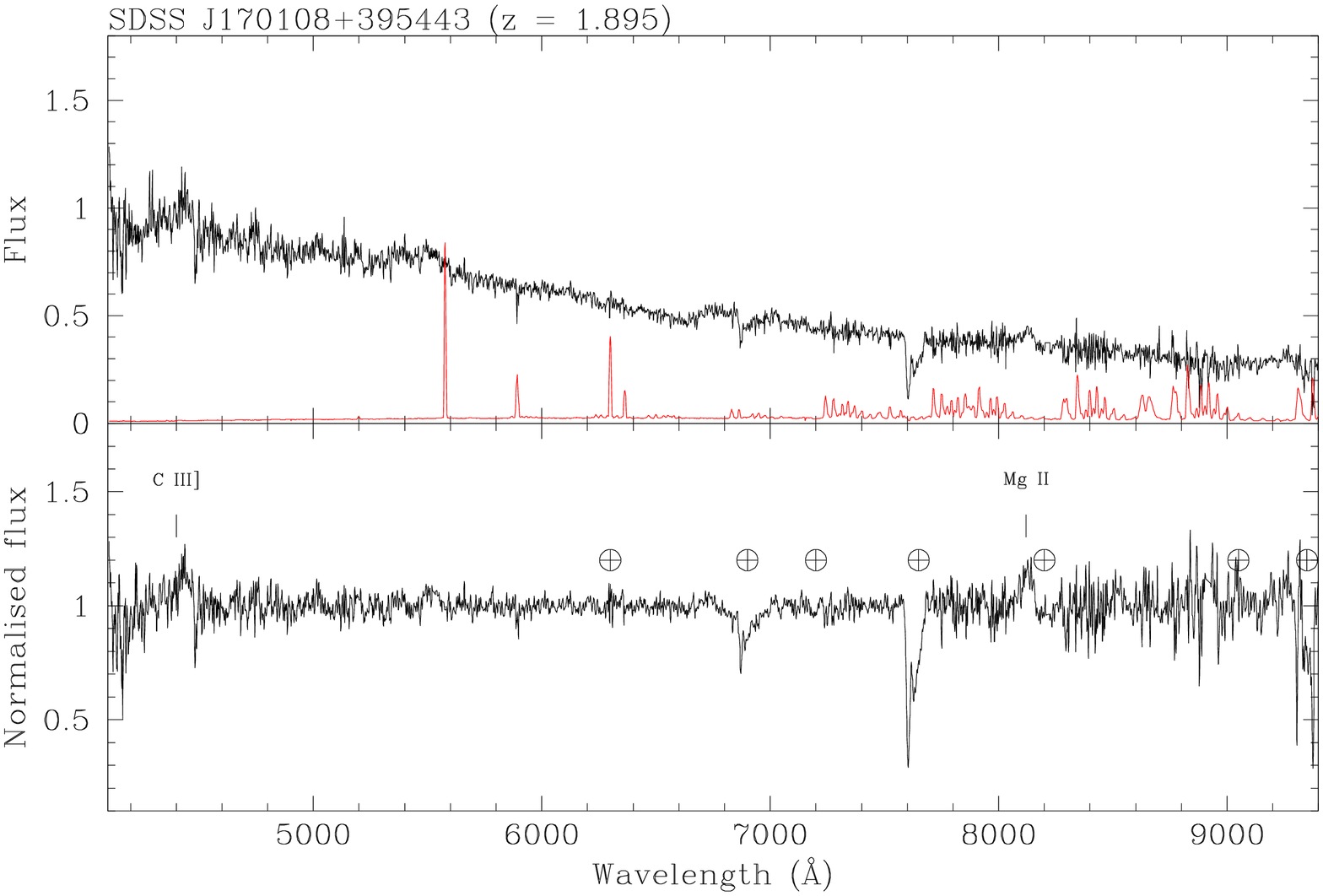}
    \label{fig:example_pair}
\end{figure*}

\newpage
\setcounter{figure}{1}
\begin{figure*}
 \includegraphics[width=0.40\textwidth, angle=-90]{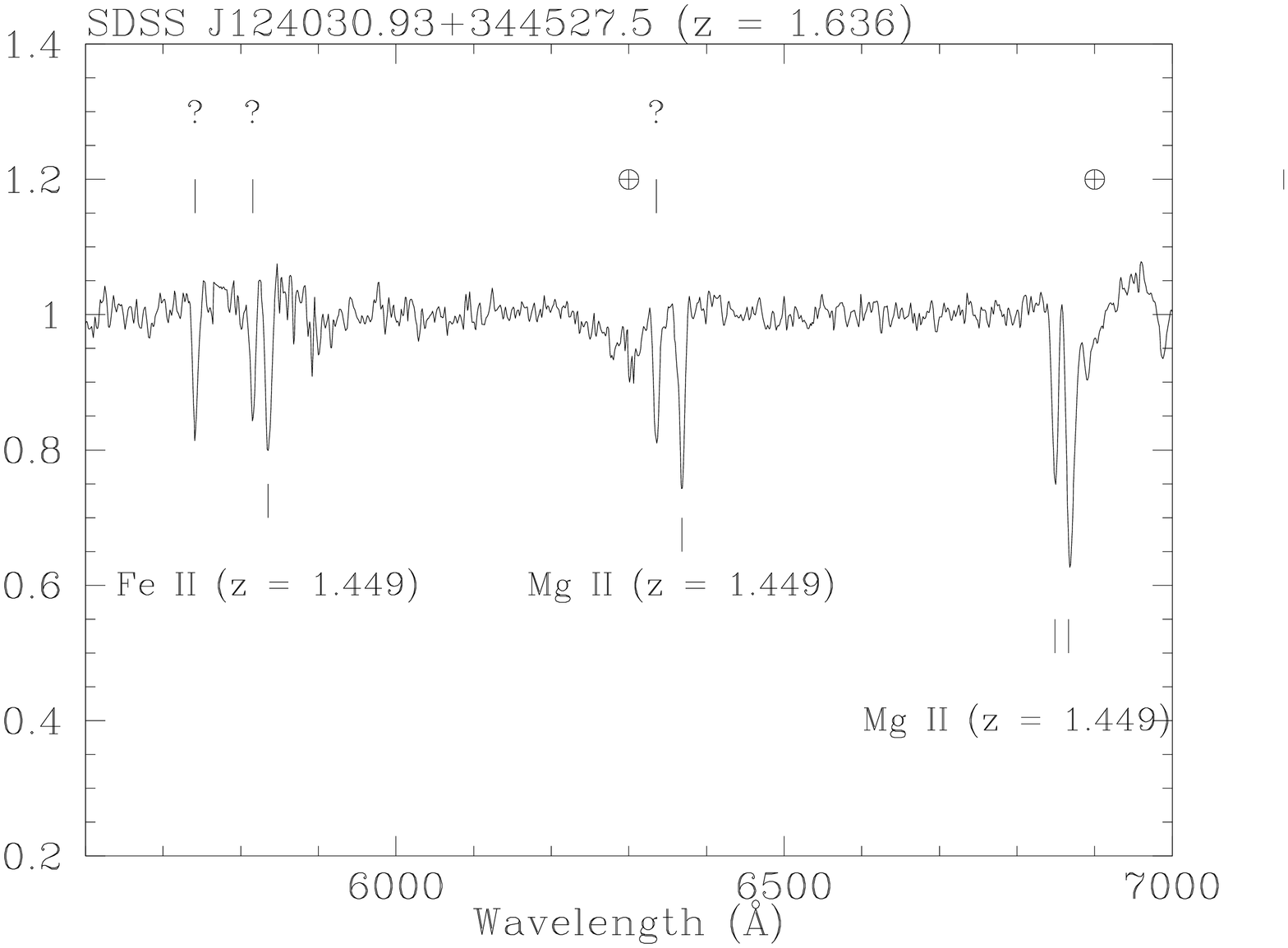}
 \includegraphics[width=0.40\textwidth, angle=-90]{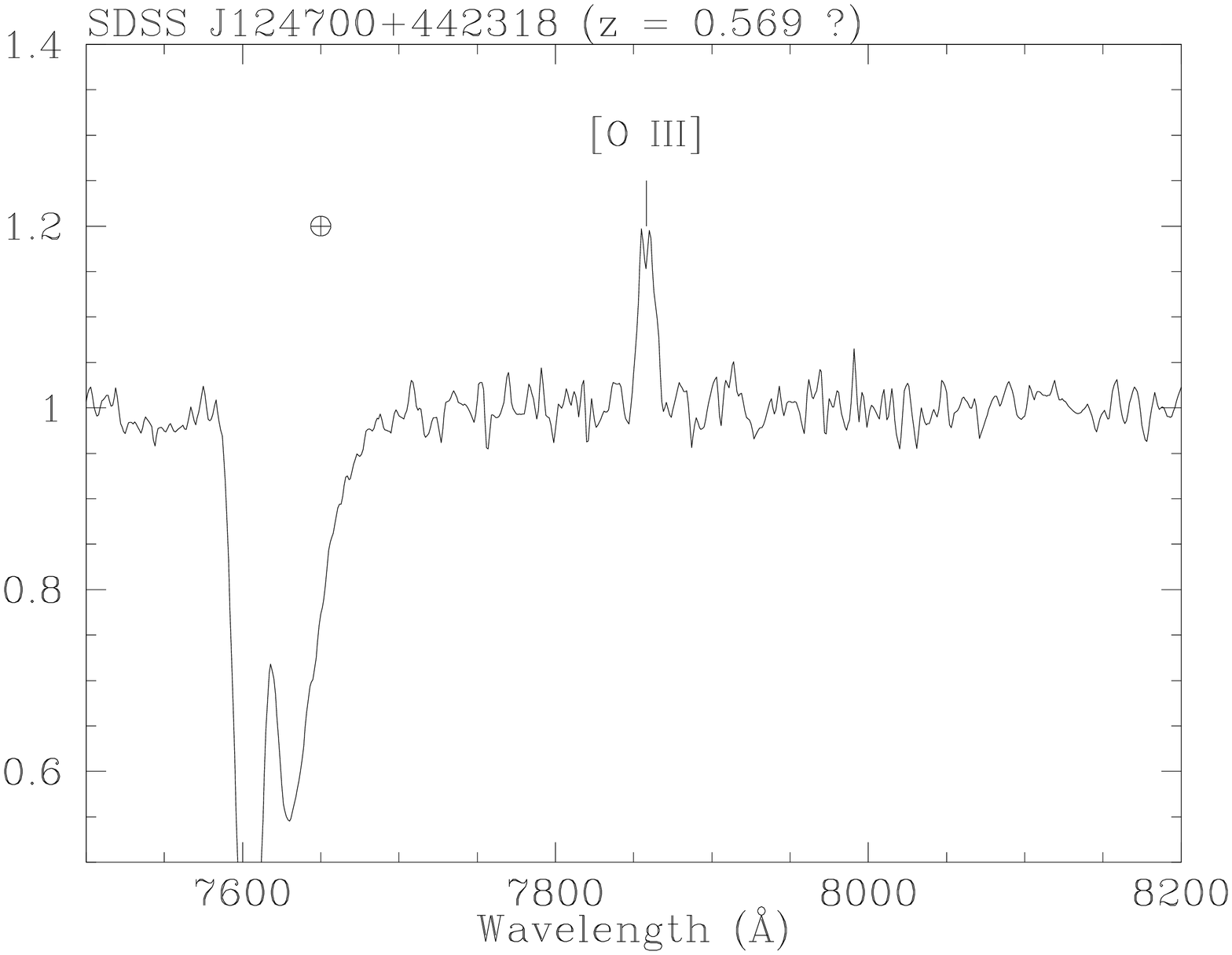}
  \includegraphics[width=0.40\textwidth, angle=-90]{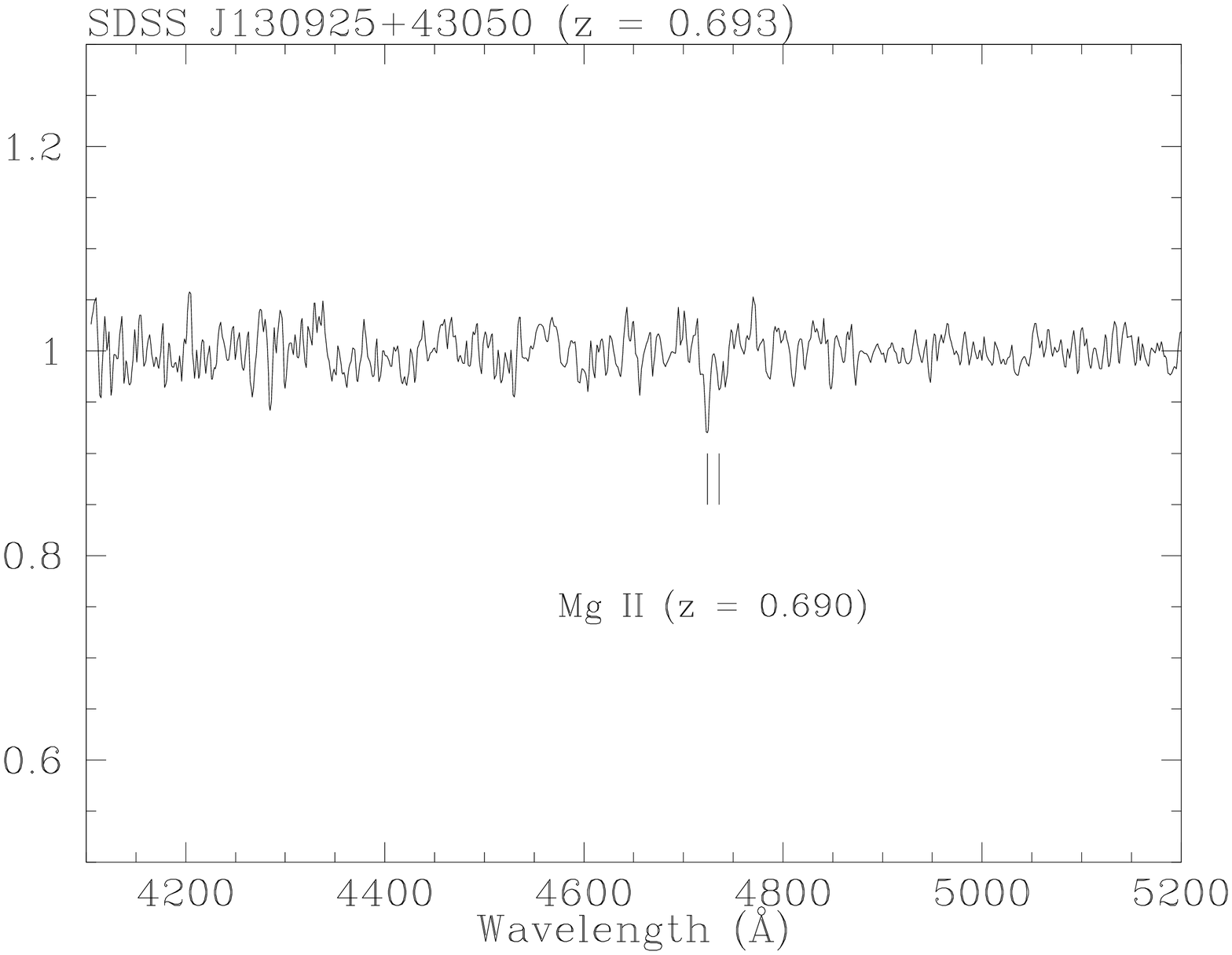} 
    \includegraphics[width=0.40\textwidth, angle=-90]{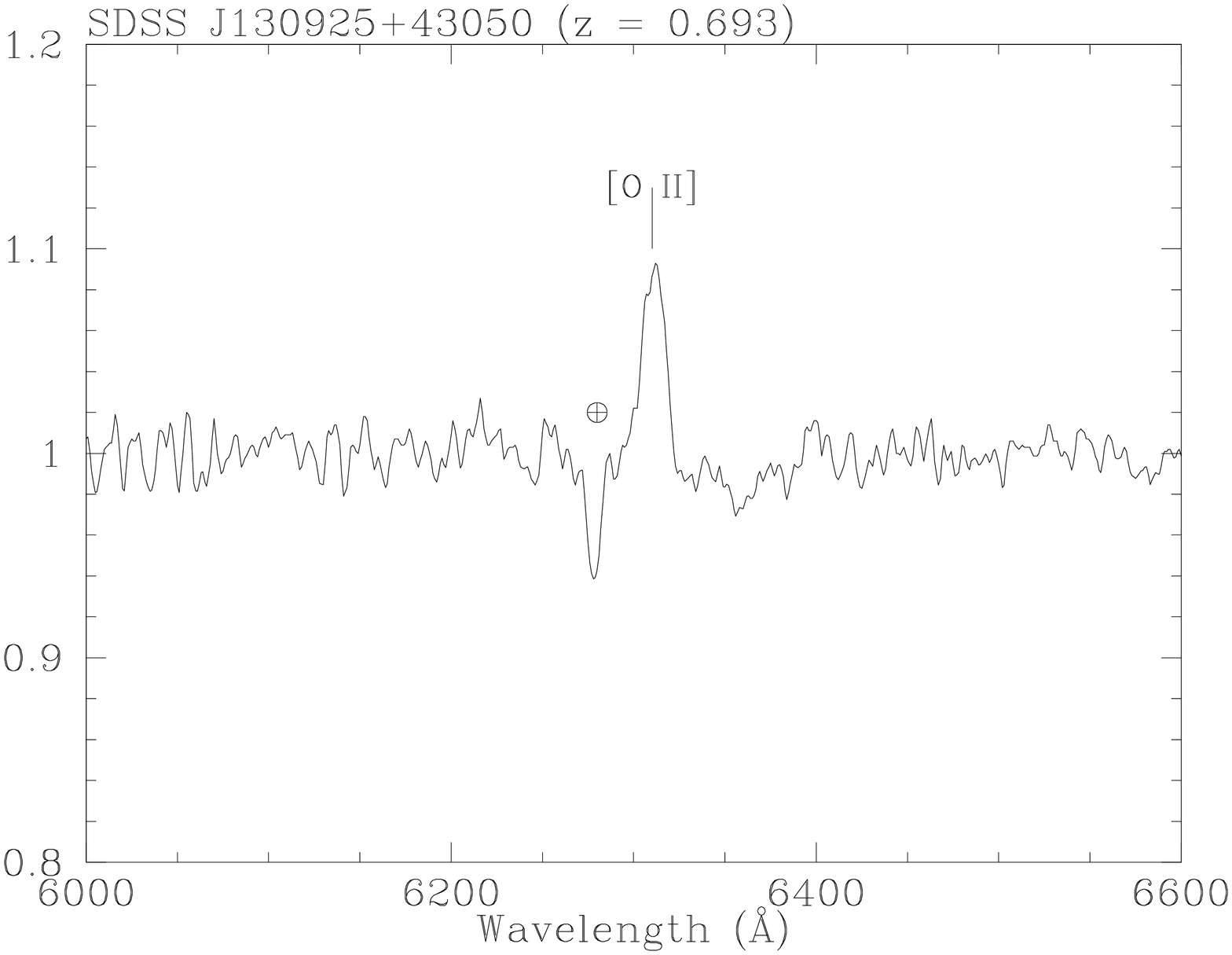} 
      \includegraphics[width=0.40\textwidth, angle=-90]{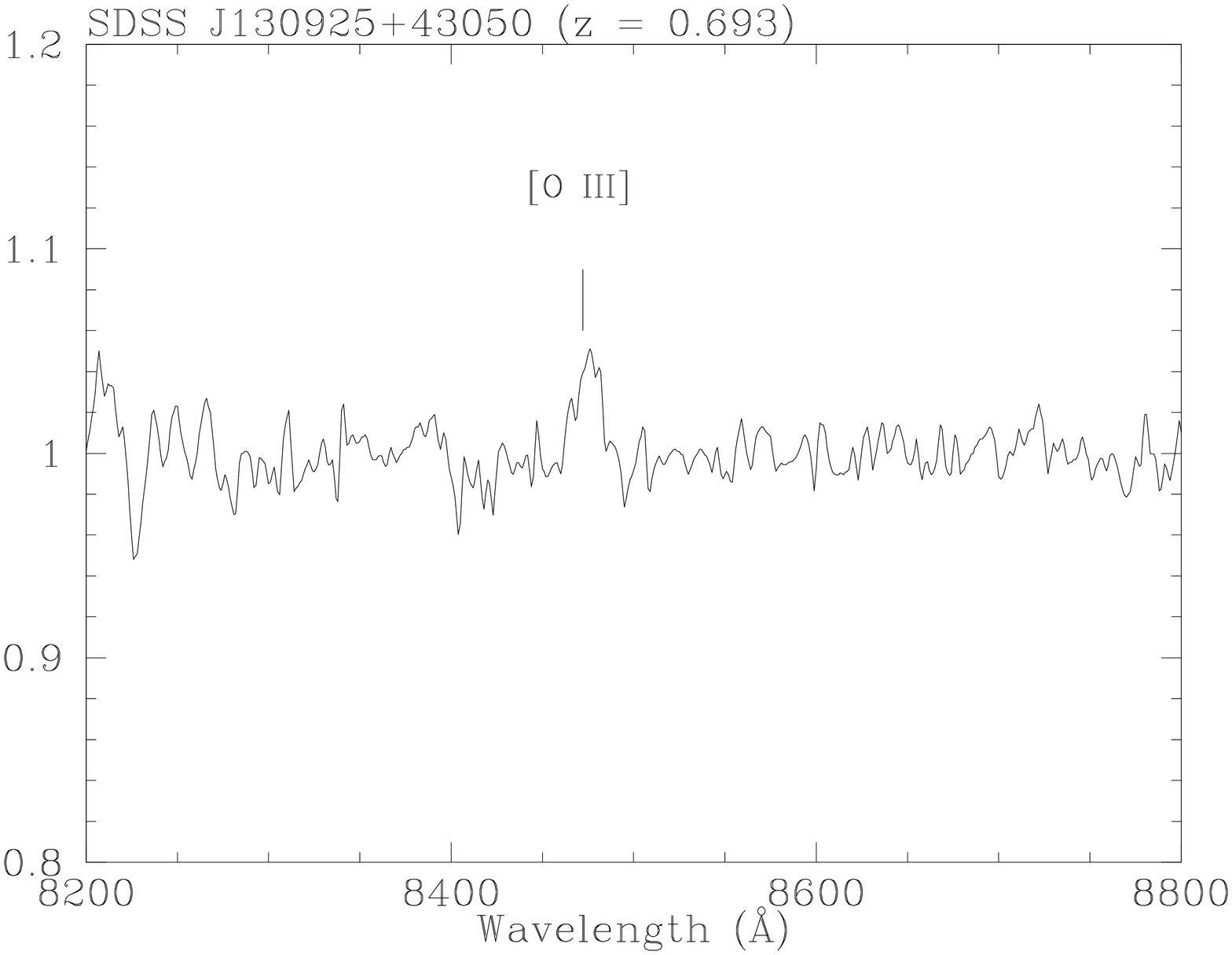} 
      \includegraphics[width=0.40\textwidth, angle=-90]{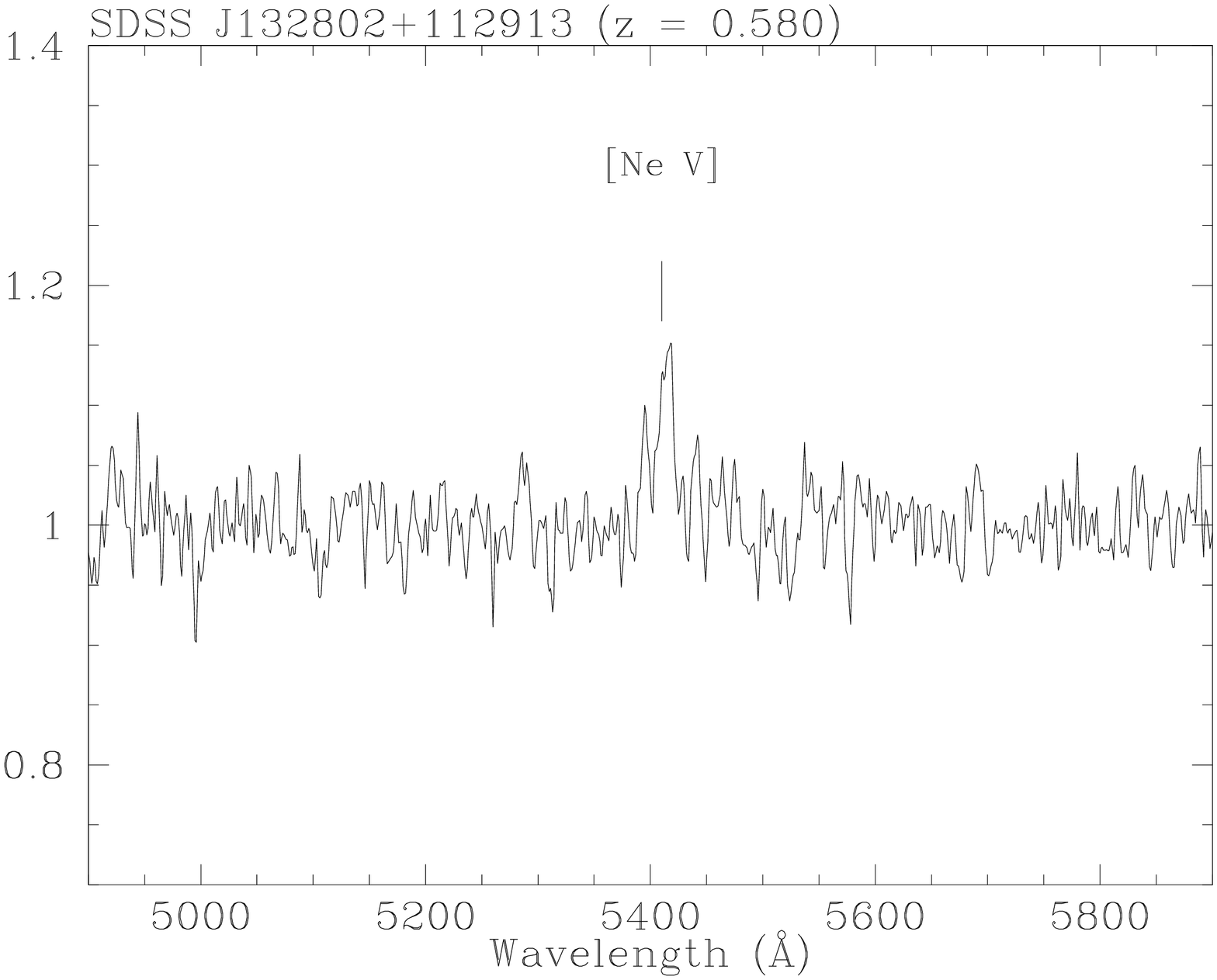} 
 \caption{Close-up of the normalized spectra around the detected spectral features. Main telluric bands are indicated as $\oplus$, spectral lines are marked by line identification. } 
   \label{fig:spectraCU}
\end{figure*}

\end{document}